\documentclass[a4paper,11pt]{article}
\usepackage{jheppub} 
\usepackage{diagbox}
\usepackage{slashbox}
\usepackage{mathtools}
\usepackage{appendix}
\usepackage[T1]{fontenc}

\newcommand{\rR}{\rho_R}
\newcommand{\gs}{g_\star}
\newcommand{\gss}{g_{\star s}}
\newcommand{\Tosc}{T_\text{osc}}
\newcommand{\Tqcd}{T_\text{QCD}}
\newcommand{\Trh}{T_\text{rh}}
\newcommand{\arh}{\mathfrak{a}_\text{rh}}
\newcommand{\aend}{\mathfrak{a}_\text{end}}
\newcommand{\rp}{\rho_\phi}

\title{
	Constraining Axion and ALP  Dark Matter from Misalignment during  Reheating}
\author[]{Yong Xu}

\affiliation[]{\it PRISMA$^+$ Cluster of Excellence and Mainz Institute for Theoretical Physics, Johannes
	Gutenberg University, 55099 Mainz, Germany}

\emailAdd{yonxu@uni-mainz.de}

\abstract{
	We explore the phenomenology of QCD axion and axion-like particle (ALP) dark matter production via misalignment during inflationary reheating. We investigate scenarios involving  inflaton oscillating in a  generic potential  $\sim \phi^n$, considering inflaton decay and annihilation for reheating. For low reheating temperatures, the parameter space leading to the correct relic abundance can be enlarged beyond the standard case. Depending on the type of inflaton-matter couplings and the value of $n$, we find that certain parts of the  extended parameter space are already constrained by  ADMX, CAPP, and MUSE experiments. Future Haloscope  experiments are expected to  impose more stringent constraints. We highlight the potential to utilize axion experiments in constraining the dynamics of reheating.
}
\begin{document}
	\begin{flushright}
		MITP-23-047
		\\October 2023   
	\end{flushright}
	\maketitle
	
	\section{Introduction}
	QCD axions and axion-like particles (ALPs) are among the well  motivated candidates for cold dark matter (DM)~\cite{Preskill:1982cy, Abbott:1982af, Dine:1982ah, Arias:2012az}. They are  pseudo-Nambu-Goldstone bosons, arising  from the spontaneous  breaking of some global $U(1)$ symmetry \cite{Peccei:1977hh, Peccei:1977ur, Weinberg:1977ma, Wilczek:1977pj} or  low energy effective field theory emerging from  string theory \cite{Conlon:2006tq, Svrcek:2006yi, Arvanitaki:2009fg, Acharya:2010zx, Cicoli:2012sz}. See Refs.~\cite{Choi:2020rgn, Agrawal:2022yvu} for reviews on the recent progress on axions and ALPs.  In the early universe, cold axion and ALP  DM can be copiously sourced\footnote{Alternatively, scatterings in the thermal plasma and the decay of topological defects can also source axion \cite{Sikivie:2006ni}. Relativistic axions can be sourced via evaporation of primordial black holes \cite{Bernal:2021yyb, Schiavone:2021imu, Bernal:2021bbv, Mazde:2022sdx}. In dense environments, such as red giants, white dwarfs or neutron stars, axions can  be produced abundantly via the Bremsstrahlung processes \cite{Turner:1989vc}.} via the vacuum misalignment mechanism~\cite{Marsh:2015xka, DiLuzio:2020wdo, Sikivie:2020zpn}. 
	
	In the standard scenario where  oscillations begin during the radiation-dominated epoch, the relic abundance of QCD axions depends on both the mass $m_a$ and the initial misalignment angle $\theta_i$. For $\theta_i \sim \mathcal{O}(1)$\footnote{Assuming that $\theta_i$ follows a uniform distribution from $-\pi$ to $\pi$, the average is $\sqrt{\left\langle \theta_i^2 \right\rangle} = \pi/\sqrt{3}$ \cite{Kolb:1990vq}.}, a mass window in the range $(10^{-6}~\text{eV}, 10^{-5}~\text{eV})$ is necessary to match observed DM relics. For ALPs, besides the mass and initial misalignment angle, the relic abundance is also dependent on the decay constant. Similarly, the parameter space is limited unless the initial misalignment angle is not $\mathcal{O}(1)$.  In literature, it is demonstrated that for low-scale inflation, axions could follow a Bunch-Davies distribution, allowing for a smaller initial misalignment angle \cite{Takahashi:2018tdu, Graham:2018jyp}, which in turn can widen the parameter space. Additionally, models with nonzero initial velocity, namely $\dot{\theta}_i \neq 0$, could also extend the axion window~\cite{Co:2019jts, Chang:2019tvx, Barman:2021rdr}. Here, we primarily focus on the simplest and most traditional scenario with $\theta_i \sim \mathcal{O}(1)$ and assume $\dot{\theta}_i = 0$ \cite{Kolb:1990vq}. 
	In this case, it has been shown that deviations from standard cosmological histories—such as the presence of early matter or kination epochs—can significantly broaden the parameter space for both axions and ALPs~\cite{Steinhardt:1983ia, Lazarides:1990xp, Kawasaki:1995vt, Giudice:2000ex, Grin:2007yg, Visinelli:2009kt, Nelson:2018via, Visinelli:2018wza, Ramberg:2019dgi, Blinov:2019rhb, Blinov:2019jqc, Bernal:2021yyb, Bernal:2021bbv, Arias:2021rer, Arias:2022qjt}.

	In this paper, we explore the generation of axion and ALP DM via misalignment mechanism  during inflationary reheating.\footnote{This is complementary to recent investigations regarding DM production during reheating  via other mechanisms, such as freeze-out or freeze-in~\cite{Drees:2017iod, Bernal:2018qlk, Maity:2018dgy, Maity:2018exj, Bernal:2019mhf, Garcia:2020eof, Bernal:2021qrl, Calibbi:2021fld, Ahmed:2022tfm, Barman:2022tzk, Banerjee:2022fiw, Bernal:2022wck, Bhattiprolu:2022sdd, Haque:2023yra, Chowdhury:2023jft, Silva-Malpartida:2023yks, Becker:2023tvd, Gan:2023jbs}.}
	Reheating affects the parameter space compatible with the observed relic abundance in two ways. First, the entropy injection during reheating is expected to enlarge the parameter space, possibly encompassing regions accessible to current and future Haloscope experiments~\cite{ Irastorza:2018dyq, Adams:2022pbo}. It is revealed that the degree of entropy injection is intricately linked to the dynamics of reheating, depending on factors such as the shape of the inflaton potential and the nature of inflaton-matter couplings \cite{Garcia:2020wiy, Bernal:2022wck}. Moreover, when assuming misalignment during reheating, the oscillation temperature (another important factor determining the parameter space) is controlled by the underlying reheating scenarios. These connections underscore the correlation between the augmented parameter space and the underlying reheating dynamics.
	
	Here, our aim is to determine the parameter space for axion and ALP DM resulting from misalignment during reheating, while also exploring the corresponding experimental constraints from Haloscope and Telescope experiments. The primary objective of this study is to investigate the impact of reheating dynamics, particularly the shape of the inflaton potential and the type of inflaton-matter couplings, on the parameter space. Furthermore, we aim to analyze the associated experimental constraints. Additionally, we explore the potential of utilizing axion experiments to constrain reheating scenarios. To achieve this, we focus on a scenario where the inflaton oscillates around a general potential proportional to $\phi^n$ during reheating.\footnote{We assume that $n$ is an even number to ensure the existence of a minimum.} Additionally, we encompass a wide range of possibilities by considering reheating scenarios arising from both inflaton decays to scalars or fermions and annihilations to scalars.
	
	Note that the current scenario is more general compared to what has been studied in the literature. To begin with, the two special cases\footnote{For a coherent scalar field oscillating around $\sim \phi^{n}$, the equation-of-state parameter is $\omega=(n-2)/(n+2)$ \cite{Turner:1983he}. Therefore,  $\omega=0$ for $n=2$ and $\omega =1$ for $n\to \infty$.} with $n=2$ and $n \to \infty$ will recover the aforementioned scenarios where misalignment occurred during early matter and kination epochs. It is then expected that our scenario with $n=2$ and $n \to \infty$ can reproduce the recent results in the literature, such as those in Ref.~\cite{Arias:2021rer} and Ref.~\cite{Blinov:2019rhb} for axions and ALPs misalignment during early matter and kination epochs. We will go beyond these two special cases; in particular, we will provide analytical expressions for any general $n$. Most importantly, we will also investigate the effect of inflaton-matter couplings on the axion and ALP DM parameter space, which has not been explored in the literature. This is particularly interesting since it offers a potential new avenue to probe reheating using axion experiments.
	
	The rest of the paper is organized as follows. In Sec.~\ref{sec:Reheating}, we revisit reheating with a particular focus on the evolution of energy densities and temperature. In Sec.~\ref{sec:after_reheating}, we briefly review misalignment in the radiation epoch after reheating. In Sec.~\ref{sec:during_reheating}, we investigate the phenomenology and parameter space for misalignment occurring during reheating. In Sec.~\ref{sec:exp}, we present the experimental constraints and comment on how axion experiments can be used to constrain reheating. Finally, we summarize our findings in Sec.~\ref{sec:conclusion}.
	
	\section{Reheating}\label{sec:Reheating}
	After cosmic inflation ends, the inflaton starts to oscillate around the minimum of its potential, transferring energy to daughter particles that eventually thermalize and form a thermal bath~\cite{Allahverdi:2010xz, Amin:2014eta, Lozanov:2019jxc}. Here, we assume that the inflaton potential during reheating takes the form $\sim \phi^n$ with $n$ being an even integer, which could originate from $\alpha$ attractor inflation models \cite{Kallosh:2013hoa}, the Starobinsky inflation \cite{Starobinsky:1980te}, or polynomial inflation models \cite{Drees:2021wgd, Drees:2022aea, Xu:2022qpx}.
	During reheating, the inflaton and radiation energy densities (denoted as $\rp$ and $\rR$, respectively) can be tracked using the following Boltzmann equations \cite{Garcia:2020wiy, Bernal:2022wck}:
	\begin{align}
	\frac{d\rp}{dt}+\frac{6n}{n+2}H\,\rp=-\frac{2n}{n+2}\Gamma\,\rp\,; \label{rp_eom}  \\
	\frac{d\rR}{dt}+4H\,\rR=+\frac{2n}{n+2}\Gamma\,\rp \label{rR_eom} \,,
	\end{align}
	where $H= \sqrt{(\rp +\rR)/(3\,M_P^2)}$ corresponds to the Hubble parameter with  $M_P$ representing the reduced Planck mass, and $\Gamma$ denotes the inflaton decay or annihilation rates into the radiation bath. In this work, we assume that the inflaton is a gauge singlet scalar field and consider inflaton decays into a pair of $\varphi$  scalars (e.g., the Higgs field in the standard model) or vector-like fermions $\psi$ (e.g., right-handed neutrinos) or annihilation into scalars. For bosonic and fermionic decays, they can proceed via trilinear interactions $\sim \mu \phi |\varphi|^2$ and $\sim y \phi \bar{\psi} \psi$; for bosonic annihilation, the interaction can be described by $\sim \lambda \phi^2 |\varphi|^2$. With these interactions, we can then compute the inflaton energy transfer rates, which are given by:
	\begin{align} \label{eq:Gamma}
	\Gamma = 
	\begin{dcases}
	\frac{\mu_{\text{eff}}^2}{8\pi\, m_\phi} &\text{bosonic decay} \,,\\
	\frac{y_{\text{eff}}^2\, m_\phi}{8\pi} 
	&\text{fermionic decay} \,,\\
	% \frac{\rho_\phi}{m_\phi} 
	% \cdot
	\frac{\lambda_{\text{eff}}^2\,\rho_\phi}{16 \pi\,m_\phi^3} 
	&\text{bosonic annihilation} \,,
	\end{dcases}
	\end{align}
	where $\mu_{\text{eff}}\,, y_{\text{eff}}\,~\text{and}~ \lambda_{\text{eff}}$ correspond to effective couplings after averaging over inflaton oscillations \cite{Ichikawa:2008ne, Garcia:2020wiy}, and $m_\phi$ denotes the inflaton mass parameter, which is given by the second derivative of the potential $ \propto \phi^{n-2}$ \cite{Garcia:2020wiy, Bernal:2022wck}. 
	
	We define the end of reheating, or the onset of radiation epoch to be the moment when $\rp(\arh) =\rR(\arh) = 3H(\arh)^2\, M_P^2$, where $\arh$ is the  corresponding scale factor at that moment. In the regime with $\Gamma \ll H$, the solution for Eq.~\eqref{rp_eom} during reheating  can be approximated as \cite{Garcia:2020wiy, Bernal:2022wck}
	\begin{align} \label{eq:rp}
	\rp(\mathfrak{a}) \simeq \rp (\arh) \left(\frac{\arh}{\mathfrak{a}}\right)^\frac{6\, n}{2 + n}\,.
	\end{align}
	The Hubble parameter can then be expressed  as follows:
	\begin{align} \label{eq:Hubble}
	H(\mathfrak{a}) \simeq H(\arh) 
	\begin{dcases}
	\left(\frac{\arh}{\mathfrak{a}}\right)^\frac{3\, n}{n + 2} &\text{ for } \aend \leq \mathfrak{a} \leq \arh\,,\\
	\left(\frac{\arh}{\mathfrak{a}}\right)^2
	&\text{ for } \arh \leq \mathfrak{a}\,,
	\end{dcases}
	\end{align}
	where $\aend$ denotes the scale factor at the end of inflaton. The first piece of Eq.~\eqref{eq:Hubble} is obtained by using  Friedmann equation with Eq.~\eqref{eq:rp}, and the second piece corresponds to the values during the radiation epoch.
	Using Eq.~\eqref{eq:Gamma}, Eq.~\eqref{eq:rp} and  Eq.~\eqref{eq:Hubble}, one can solve  Eq.~\eqref{rR_eom} and obtain the solution for $\rR$ during reheating \cite{Garcia:2020wiy, Bernal:2022wck}:
	\begin{align}\label{eq:rR_sol}
	\rR(\mathfrak{a}) \simeq 
	\begin{dcases}
	\frac{n }{1+2n}\, \rp(\arh) \left(\frac{\arh}{\mathfrak{a}}\right)^{\frac{6}{n+2}}\left[1-\left(\frac{\aend}{\mathfrak{a}}\right)^{\frac{2+4n}{2+n}}\right]    &\text{bosonic decay,}\\
	\frac{n }{7-n}\, \rp(\arh) \left(\frac{\arh}{\mathfrak{a}}\right)^{4}\left[\left(\frac{\aend}{\mathfrak{a}}\right)^{\frac{2(n-7)}{2+n}}-1\right]  &\text{fermionic decay,}\\
	\frac{n }{2n-5}\, \rp(\arh) \left(\frac{\arh}{\mathfrak{a}}\right)^{\frac{18}{n+2}}\left[1-\left(\frac{\aend}{\mathfrak{a}}\right)^{\frac{2(2n-5)}{2+n}}\right]  &\text{bosonic annihilation.}\\
	\end{dcases}
	\end{align}
	From Eq.~\eqref{eq:rR_sol}, we can further compute the thermal bath temperature $T (\mathfrak{a})= \left[\rR(\mathfrak{a}) \, 30  /(\pi^2 \gs )\right]^{1/4} $, where $\gs$ denotes the degrees of freedom contributing the radiation energy densities. We find that the thermal bath temperature can be expressed as the following general expression \cite{Garcia:2020wiy, Bernal:2022wck}:
	\begin{align} \label{eq:Tevol}
	T(\mathfrak{a}) \simeq \Trh 
	\begin{dcases}
	\left(\frac{\arh}{\mathfrak{a}}\right)^\alpha &\text{ for }  \aend \leq \mathfrak{a} \leq \arh \,,\\
	\left(\frac{\arh}{ \mathfrak{a}}\right)^1
	&\text{ for } \arh \leq \mathfrak{a}\,,
	\end{dcases}
	\end{align}
	where the $\alpha$ parameters  are given by  
	\begin{align}
	\alpha = 
	\begin{dcases}
	\frac{3}{2(n+2)} &\text{bosonic decay,}\\ 
	\frac{3(n-1)}{2(n+2)}   &\text{fermionic decay,}\\
	\frac{9}{2(n+2)}  &\text{bosonic annihilation.}
	\end{dcases}
	\end{align}
	We can then write the Hubble parameter in Eq.~\eqref{eq:Hubble} as function of temperature:
	\begin{align} \label{eq:HubbleT}
	H(T) \simeq 
	\begin{dcases}
	H(\Trh) \left(\frac{T}{\Trh}\right)^{\frac{3\, n}{2 + n}\, \frac{1}{\alpha}} &\text{ for } \Trh \leq T  \,,\\
	H(\Trh) \left(\frac{T}{\Trh}\right)^{2} &\text{ for } T \leq \Trh\,.
	\end{dcases}
	\end{align}

	Several comments in order before closing this section. First,  for bosonic annihilation with $n=2$, it is not possible for the radiation to surpass inflaton energy densities as $\rR \propto (\arh/\mathfrak{a})^4$ while $\rp \propto (\arh/\mathfrak{a})^3$ during reheating. 
	Therefore, we need to consider  $n>2$ to achieve successful reheating and allow the universe to transition into a radiation-dominated epoch.
	Secondly, in the fermionic decay scenario with for $n>7$, one has $\alpha=1$  because $\left(\aend/\mathfrak{a}\right)^{2(n-7)/(2+n)} \ll 1$ (in the second line of Eq.~\eqref{eq:rR_sol}).  Furthermore, the analytical results, in particular Eq.~\eqref{eq:Tevol} and Eq.~\eqref{eq:HubbleT}, rely on the assumption that the dominant contribution for estimating temperature arises from perturbative inflaton decays. This assumption remains valid under the condition that the impact of non-perturbative phenomena \cite{Lozanov:2019jxc, Lozanov:2016hid}, and gravitational effects during reheating \cite{Haque:2022kez} remain subordinate to perturbative decay. It is important to note that even though  the presence of non-perturbative processes could lead to the equation-of-state parameter $\omega$ approaching $1/3$, perturbative decay retains its significance in fully depleting inflaton energy \cite{Dufaux:2006ee, Maity:2018qhi, Saha:2020bis}. Moreover, for the bosonic decay scenario, investigations have highlighted the inefficacy of preheating due to the backreaction stemming from the self-interaction of the daughter field \cite{Dufaux:2006ee}. Similarly, fermionic preheating could be impeded by Pauli blocking \cite{Peloso:2000hy}. These conclusions are expected to be more robust for weak couplings between the inflaton and daughter particles, a regime that aligns with low reheating temperatures. Finally, the gravitational reheating mechanism demands $\omega = (n-2)/(n+2) \gtrsim 0.65$ to exert efficiency \cite{Haque:2022kez,Co:2022bgh, Haque:2023yra, Barman:2022qgt}, which effectively confines its significance to cases where $n \lesssim 9$.
	
	In this study, we mainly focus on $2 \leq n \leq 8$ with low reheating temperatures characterized by weak couplings, specifically $\Trh \lesssim 1~\text{GeV}$. In these situations, it is reasonable to concentrate on perturbative processes and utilize the elementary theory of reheating based on perturbative decay \cite{Kolb:1990vq}.
	%%%%%%%%%%%%%%%%%%%%%%%%%%%%%%%%%%%%%%%%%%%%%%%%%%%%%%%%%%
	\section{Misalignment after Reheating: $\Trh \geq \Tosc$} \label{sec:after_reheating}
	%%%%%%%%%%%%%%%%%%%%%%%%%%%%%%%%%%%%%%%%%%%%%%%%%%%%%%%%%%
	In this section we revisit the standard case where axion and ALP misalignment occurs during radiation epoch \textit{after} reheating. 
	\subsection{QCD Axion}
	The axion mass $m_a$ at zero temperature is given by~\cite{GrillidiCortona:2015jxo}
	\begin{align}\label{eq:ma1}
	m_a \simeq 5.7 \cdot 10^{-6} \left(\frac{10^{12}~\text{GeV}}{f_a}\right) \text{eV}\,,
	\end{align}
	where $f_a$ denotes the decay constant. The temperature-dependent axion mass $\tilde m_a$ was numerically calculated in Ref.~\cite{Borsanyi:2016ksw}, which can be analytically approximated  as~\cite{Arias:2021rer, Arias:2022qjt}
	\begin{align}\label{eq:ma2}
	\tilde m_a(T) \simeq m_a 
	\begin{dcases}
	(\Tqcd/T)^4 & \text{for } T \geq \Tqcd\,,\\
	1 & \text{for } T \leq \Tqcd\,,
	\end{dcases}
	\end{align}
	where $\Tqcd \simeq 150$~MeV. 

	The  Lagrangian density for axion field  is
	\begin{align}
	\mathcal{L}_a \supset \frac{1}{2}\partial^\mu a\, \partial_\nu a - \tilde{m}^2_a(T) f_a^2 \left[1-\left(\cos\frac{a}{f_a}\right)\right]\,,
	\end{align}
	with which one can derive the equation of motion of the axion field:
	\begin{align}
	\ddot{\theta} + 3 H\, \dot{\theta} + \tilde{m}^2_a(T)\, \sin \theta =0\,,
	\end{align}
	where $H$ denotes the Hubble expansion rate and  $\theta \equiv a(t)/f_a$. At  high temperature with $T\gg \Tqcd$, the Hubble parameter is much larger than the axion mass so that  the axion field is frozen to be constant. Axions begin to oscillate at the temperature $T = \Tosc$ defined by $3\,H(\Tosc) \equiv \tilde m_a(\Tosc)$ \cite{Kolb:1990vq}. In the radiation epoch, the corresponding oscillation temperature  is
	\begin{align}\label{eq:Tosc_RD}
	\Tosc \simeq 
	\begin{dcases} 
	\left(\frac{1}{\pi}\sqrt{\frac{10}{\gs}}\,m_a\, M_P \right)^{1/2} &  \Tosc \leq \Tqcd \,,\\
	\left(\frac{1}{\pi} \sqrt{\frac{10}{\gs}}\,m_a\, M_P\, \Tqcd^4\right)^{1/6}   & \Tosc \geq  \Tqcd,
	\end{dcases}
	\end{align}
	where we have used the second line of 
	Eq.~\eqref{eq:HubbleT} for the Hubble parameter.
	Under the assumption that the  entropy is conserved after reheating, we can relate the axion energy density at present with that at oscillation:
	\begin{align} \label{rho0}
	\rho_a(T_0) = \rho_a(\Tosc) \frac{m_a}{\tilde m_a(\Tosc)} \frac{s(T_0)}{s(\Tosc)}\,,
	\end{align}
	where $T_0 \simeq 2.73 ~K$ denotes the temperature  at present, and $\rho_a(\Tosc) \simeq \frac12 \tilde m_a^2(\Tosc)\, f_a^2\, \theta_i^2$ with $\theta_i$ being the initial misalignment angle. The entropy density is defined as  
	\begin{align}
	s(T) = \frac{2\pi^2}{45}\, \gss(T)\, T^3\,,
	\end{align}
	where $\gss$ denotes the degrees of freedom contributing to the SM entropy.  
	Using Eq.~\eqref{rho0} and Eq.~\eqref{eq:ma2}, we can  compute the axion relic abundance, which is
	\begin{align} \label{eq:Omega_RD}
	\Omega_ah^2 & \equiv \frac{\rho_a(T_0)}{\rho_c/h^2} \simeq 0.12 \left(\frac{\theta_i}{1.0}\right)^2 
	\begin{dcases}
	\left(\frac{m_a}{5.2 \cdot 10^{-7}~\text{eV}}\right)^{-\frac32} & \text{for } m_a \leq m_a^\text{QCD},\\
	\left(\frac{m_a}{8.5 \cdot 10^{-6}~\text{eV}}\right)^{-\frac76} & \text{for } m_a \geq m_a^\text{QCD},
	\end{dcases}
	\end{align}
	with $\rho_c =1.0 5 \cdot 10^{-5}~h^2~\text{GeV}^3/\text{cm}^3$  being the critical energy density, $s(T_0) \simeq 2.69 \cdot 10^3$~cm$^{-3}$~\cite{ParticleDataGroup:2022pth}, and  $m_a^\text{QCD} \equiv m_a(\Tosc=\Tqcd) =3 H(\Tqcd) \simeq 4.8 \cdot 10^{-11}$~eV. By assuming an  initial misalignment angle $\theta_i \simeq 1.0$,  we see from Eq.~\eqref{eq:Omega_RD} that a axion mass around $m_a \sim  \mathcal{O}(10^{-5})~\text{eV}$, and correspondingly a decay constant $f_a \sim \mathcal{O}\left(10^{12}\right)~\text{GeV}$ is required in order to match the observed relic abundance. 
	
	\subsection{ALP}
	For ALP, we remain model agnostic about the origin of  its mass. We consider ALP mass to be time independent, and the oscillation temperature is then given by the first line of Eq.~\eqref{eq:Tosc_RD}. The energy density at present takes a form:
	\begin{align} \label{eq:_rhoALP1}
	\rho_a(T_0) = \rho_a(\Tosc)\, \frac{s(T_0)}{s(\Tosc)}\,,
	\end{align}
	where $\rho_a(\Tosc) \simeq \frac12 f_a^2  m_a^2\,  \theta_i^2$. Similar as before, we can further compute the relic abundance for ALPs, which is
	\begin{align} \label{eq:Omega_a_case2}
	\Omega_ah^2 \simeq 0.12\left(\frac{\theta_i}{1.0}\right)^2  \left(\frac{f_a}{1.0 \cdot 10^{14}~\text{GeV}}\right)^2 \left(\frac{m_a}{ 7.4\cdot 10^{-11}~\text{eV}}\right)^{1/2}\,.
	\end{align}
	Note that in general $f_a$ and $m_a$ are independent parameters for ALP, however it is required that $f_a \propto m_a^{-1/4} $ in order to match the observed DM relic abundance. Besides, as a way of cross checking, we notice that once replacing $f_a$ to be $m_a$ via Eq.~\eqref{eq:ma1}, Eq.~\eqref{eq:Omega_a_case2} reproduces the first line of Eq.~\eqref{eq:Omega_RD}.
	
	In the next section, we will explore the phenomenological consequence of misalignment occurring {\it during } reheating. Depending on the dynamics of reheating, we will show that both the axion and ALP DM parameter space can be enlarged and show distinct behaviours. 
	
	\section{Misalignment during Reheating: $\Trh <\Tosc$}\label{sec:during_reheating}
	In the previous section we have assumed that the oscillation temperature is $\Tosc< \Trh$ so that misalignment occurs in radiation epoch {\it after} reheating. 
	In this section we assume axion or ALP oscillation {\it during} reheating with $\Tosc > \Trh$. In such case,  
	the  density at present is
	\begin{align} \label{eq:rhoa}
	\rho_a(T_0) = \rho_a(\Tosc) \frac{m_a}{\tilde m_a(\Tosc)} \frac{s(T_0)}{s(\Tosc)}  \frac{S(\Tosc)}{S(\Trh)}\,,
	\end{align}
	where the entropy dilution factor is given by
	\begin{align}\label{eq:dilution_S}
	\frac{S(T)}{S(\Trh)} 
	& = \left( \frac{ \gss(T) }{\gss(\Trh)}  \right) \left( \frac{T}{\Trh}   \right)^3  \left( \frac{\mathfrak{a}(T)}{\arh}  \right)^3\nonumber= \left( \frac{ \gss(T) }{\gss(\Trh)}  \right) \left( \frac{T}{\Trh}   \right)^{\frac{3\alpha - 3}{\alpha}}  \nonumber\\
	&=\left( \frac{ \gss(T) }{\gss(\Trh)}  \right)
	\begin{dcases}
	\left( \frac{\Trh}{T}   \right)^{2n+1}  &  \text{bosonic decay} \,, \\
	\left( \frac{\Trh}{T}   \right)^{\frac{7-n}{n-1}} & \text{fermionic decay}   \,, \\
	\left( \frac{\Trh}{T}   \right)^{\frac{2n-5}{3}} & \text{bosonic annihilation}  \,.
	\end{dcases}
	\end{align}
	The remaining task is to work out the oscillation temperature $\Tosc$ for a given $\alpha$ parameter, or a type of reheating scenario.
	% For the  oscillation  to occur {\it during} reheating, it is required that $\Trh< \Tosc $.
	Note that for fermionic decay with $n>7$, one has $\alpha=1$ as argued earlier, leading to $S(T)/S(\Trh)  =  \gss(T) / \gss(\Trh) $. 

	\subsection{QCD Axion}
	For QCD axion,  there are two possibilities: 
	Case 1: $ \Tqcd  < \Tosc$ and Case 2: $ \Tqcd  > \Tosc$.
	In the following, we investigate these cases separately.
	
	\subsubsection{Case 1: $ \Tqcd  < \Tosc$ }\label{sec:QCD_case1}
	In the case with  $ \Tqcd  < \Tosc$, axion mass features a temperature dependence: $\tilde{m}_a =m_a (T/\Tqcd^4)$. Using the first line of Eq.~\eqref{eq:HubbleT}, we work out the oscillation temperature, which is given by
	\begin{align}\label{eq:Tosc_case2}
	\Tosc = \Trh 
	\left(\frac{1}{\pi}\sqrt{\frac{10}{\gs}} \frac{m_a\,M_P\, \Tqcd^4}{\Trh^6}\right)^{\frac{\alpha (2+n)}{3n + 4\alpha(2 + n)}} \,.
	\end{align}
	Depending on the hierarchy of the QCD scale and the reheating temperature, there are further two possibilities: $\Trh < \Tqcd  <\Tosc$  and  $\Tqcd <\Trh  < \Tosc$, which lead to constraints on reheating temperature; for the former, we have:
	\begin{align}
	\Tqcd^{\frac{3n}{3n-2\alpha(n+2)}} \left(\frac{1}{\pi} \sqrt{\frac{10}{\gs}} m_a\,M_P\right)^{\frac{-\alpha(n+2)}{3n-2\alpha(n+2)}} < \Trh <   \Tqcd\,,
	\end{align}
	while the latter requires
	\begin{align}
	\Tqcd < \Trh  <\left(\frac{1}{\pi} \sqrt{\frac{10}{\gs}} m_a\,M_P \Tqcd^4 \right)^{\frac{1}{6}}\,.
	\end{align}
	With the oscillation temperature Eq.~\eqref{eq:Tosc_case2} and entropy dilution factor Eq.~\eqref{eq:dilution_S}, we can further proceed to compute the   axion density Eq.~\eqref{eq:rhoa}, and finally the relic abundance at present.
	
	For bosonic decay, we find a general  expression in the following form\footnote{The exponential function  ``$\text{exp}$'' has been introduced to make the result more compact.}:
	\begin{align}
	\Omega_ah^2 \simeq  0.12  \left(\frac{\theta}{1.0}\right)^{2}    \left( \frac{\Trh}{0.1~\text{GeV}}\right)^{\frac{10-n}{n+2}} 
	\left[ \frac{m_a}{2.59\cdot 10^{-6}\, \cdot \text{exp}\left(\frac{-20.19 + 3.68\, n}{n+4}\right)~\text{eV}}\right]^{-\frac{4+n}{n+2}}\,,
	\end{align}
	or with specific values of $n$
	\begin{align}
	\Omega_ah^2 \simeq  0.12  \left(\frac{\theta}{1.0}\right)^{2}
	\begin{dcases}
	\left( \frac{\Trh}{0.1~\text{GeV}}\right)^{2}  \left( \frac{m_a}{3.1 \cdot 10^{-7}~\text{eV} }\right)^{-\frac{3}{2}}& \text{bosonic decay $n=2$,} \\
	\left( \frac{\Trh}{0.1 ~\text{GeV}}\right)^{1}  \left( \frac{m_a}{1.3 \cdot 10^{-6}~\text{eV} }\right)^{-\frac{4}{3}}& \text{bosonic decay $n=4$,} \\
	\left( \frac{\Trh}{0.1~\text{GeV}}\right)^{\frac{1}{2}}  \left( \frac{m_a}{3.1 \cdot 10^{-6}~\text{eV} }\right)^{-\frac{5}{4}} & \text{bosonic decay $n=6$,} \\
	\left( \frac{\Trh}{0.1~\text{GeV}}\right)^{\frac{1}{5}}  \left( \frac{m_a}{5.6 \cdot 10^{-6}~\text{eV} }\right)^{-\frac{6}{5}} & \text{bosonic decay $n=8$.} \\
	\end{dcases}
	\end{align}
	Similarly,  for fermionic decay, we find 
	\begin{align}
	\Omega_ah^2 \simeq  0.12  \left(\frac{\theta}{1.0}\right)^{2} 
	\begin{dcases}
	\left( \frac{\Trh}{0.1~\text{GeV}}\right)^{\frac{14-3n}{3n-2}}  \left[ \frac{m_a}{1.03\cdot 10^{-6}\cdot \text{exp} \left(\frac{-11.64 + 4.61\, n}{n+2}\right) ~\text{eV} }\right]^{\frac{3n}{2-3n}} & n<7\,, \\
	\left( \frac{\Trh}{0.1~\text{GeV}}\right)^{\frac{7(4-n)}{8+7n}}  \left[ \frac{m_a}{1.03\cdot 10^{-6}\cdot \text{exp} \left(\frac{-5.76 + 4.61\, n}{n+2}\right) ~\text{eV} }\right]^{\frac{-7(2+n)}{8+7n}} & n>7\,,
	\end{dcases}
	\end{align}
	and 
	\begin{align}
	\Omega_ah^2 \simeq  0.12  \left(\frac{\theta}{1.0}\right)^{2}
	\begin{dcases}
	\left( \frac{\Trh}{0.1~\text{GeV}}\right)^{2}  \left( \frac{m_a}{3.1 \cdot 10^{-7}~\text{eV} }\right)^{-\frac{3}{2}}& \text{fermionic decay $n=2$,} \\
	\left( \frac{\Trh}{0.1 ~\text{GeV}}\right)^{\frac{1}{5}}  \left( \frac{m_a}{5.6 \cdot 10^{-6}~\text{eV} }\right)^{-\frac{6}{5}}& \text{fermionic decay $n=4$,} \\
	\left( \frac{\Trh}{0.1~\text{GeV}}\right)^{-\frac{1}{4}}  \left( \frac{m_a}{1.5 \cdot 10^{-5}~\text{eV} }\right)^{-\frac{9}{8}} & \text{fermionic decay $n=6$,} \\
	\left( \frac{\Trh}{0.1~\text{GeV}}\right)^{-\frac{7}{16}}  \left( \frac{m_a}{2.3 \cdot 10^{-5}~\text{eV} }\right)^{-\frac{35}{32}} & \text{fermionic decay $n=8$.} \\
	\end{dcases}
	\end{align}
	For bosonic annihilation, we have 
	\begin{align}
	\Omega_ah^2 \simeq  0.12  \left(\frac{\theta}{1.0}\right)^{2} \left( \frac{\Trh}{0.1~\text{GeV}}\right)^{\frac{6-n}{6+n}}  \left[ \frac{m_a}{1.0\cdot 10^{-6}\cdot \text{exp} \left(\frac{2.14+4.63\, n}{n+8}\right) ~\text{eV} }\right]^{\frac{-(8+n)}{6+n}}
	\end{align}
	and
	\begin{align}
	\Omega_ah^2 \simeq  0.12  \left(\frac{\theta}{1.0}\right)^{2}
	\begin{dcases}
	\left( \frac{\Trh}{0.1 ~\text{GeV}}\right)^{\frac{1}{5}}  \left( \frac{m_a}{5.6 \cdot 10^{-6}~\text{eV} }\right)^{-\frac{6}{5}}& \text{bosonic annihilation $n=4$,} \\
	\left( \frac{\Trh}{0.1~\text{GeV}}\right)^{0}  \left( \frac{m_a}{8.5 \cdot 10^{-6}~\text{eV} }\right)^{-\frac{7}{6}} & \text{bosonic annihilation $n=6$,} \\
	\left( \frac{\Trh}{0.1~\text{GeV}}\right)^{-\frac{1}{7}}  \left( \frac{m_a}{1.2 \cdot 10^{-5}~\text{eV} }\right)^{-\frac{8}{7}} & \text{bosonic annihilation $n=8$.} \\
	\end{dcases}
	\end{align}

	For the scenario with $n=2$, we observe that the expression for $\Omega_ah^2$ remains the same for both the bosonic and fermionic decay reheating scenarios. This similarity arises because, in this case, the inflaton decay rate $\Gamma$ does not exhibit time dependence due to the constant inflaton mass. Consequently, both the dilution factor Eq.~\eqref{eq:dilution_S} and the oscillation temperature Eq.~\eqref{eq:Tosc_case2} are identical for both bosonic and fermionic decay reheating scenarios with $n=2$. Additionally, we notice that in the case of bosonic annihilation with $n=6$, the relic abundance becomes independent of the reheating temperature $\Trh$, remaining the same as in the case of oscillations occurring during the radiation-dominated epoch, as shown in the second line of Eq.~\eqref{eq:Omega_RD}. 
	
	\subsubsection{Case 2: $ \Tqcd  \geq \Tosc$ } \label{sec:QCD_case2}
	For oscillation temperature below QCD scale, we have $\tilde{m}_a =m_a$. Using the first line of Eq.~\eqref{eq:HubbleT}, we find the oscillation temperature 
	\begin{align}\label{eq:Tosc}
	\Tosc = \Trh 
	% \left(\right)^{\frac{\alpha (2+n)}{3n}}  
	\left(\frac{1}{\pi}\sqrt{\frac{10}{\gs}} \frac{m_a \, M_P}{\Trh^2}\right)^{\frac{\alpha (2+n)}{3n}} \,.
	\end{align}
	Similar as before, the self-consistency condition $\Trh \leq \Tosc \leq \Tqcd$  yields bounds on reheating temperature, which in this case lead to
	\begin{align}
	& \Trh \leq  \Tqcd^{\frac{3n}{3n-2\alpha(2+n)}}    \left(\frac{1}{\pi} \sqrt{\frac{10}{\gs}} m_a\,M_P\right)^{\frac{-\alpha(2+n)}{3n-2\alpha(2+n)}}\,;  \label{eq:Trhbound}\\
	&\Trh \leq \left(\frac{1}{\pi} \sqrt{\frac{10}{\gs}} m_a\,M_P\right)^{\frac{1}{2}} \label{eq:Trhbound2}\,.
	\end{align}

	With the oscillation temperature, we find the relic abundance reads
	\begin{align}\label{eq:Omega_a}
	\Omega_ah^2 \simeq  0.12  \left(\frac{\theta}{1.0}\right)^{2}   \left( \frac{\Trh}{4 ~\text{MeV}}\right)^{\frac{4-n}{n}}   \left[\frac{m_a}{1.0\cdot 10^{-6}\cdot \text{exp}\left(\frac{-35.29 + 7.85~ n}{n+2}\right)~\text{eV} }\right]^{-\frac{2+n}{n}}\,,
	\end{align}
	or with specific value of $n$:
	\begin{align} \label{eq:Omega_a_n}
	\Omega_ah^2 \simeq  0.12  \left(\frac{\theta}{1.0}\right)^{2}
	\begin{dcases}
	\left( \frac{\Trh}{4~\text{MeV}}\right)^{1}  \left( \frac{m_a}{7.5\cdot 10^{-9}~\text{eV} }\right)^{-2}& \text{$n=2$,} \\
	\left( \frac{\Trh}{4 ~\text{MeV}}\right)^{0}  \left( \frac{m_a}{5.2 \cdot 10^{-7}~\text{eV} }\right)^{-\frac{3}{2}}& \text{$n=4$,} \\
	\left( \frac{\Trh}{4~\text{MeV}}\right)^{-\frac{1}{3}}  \left( \frac{m_a}{4.4 \cdot 10^{-6}~\text{eV} }\right)^{-\frac{4}{3}} & \text{$n=6,$} \\
	\left( \frac{\Trh}{4~\text{MeV}}\right)^{-\frac{1}{2}}  \left( \frac{m_a}{1.6 \cdot 10^{-5}~\text{eV} }\right)^{-\frac{5}{4}} & \text{$n=8.$} \\
	\end{dcases}
	\end{align}
	It is interesting to note that the relic abundance is independent of the $\alpha$  parameter or the form of inflaton-matter couplings, even though the oscillation temperature still depends on them. This becomes clear to recall from Eq.~\eqref{eq:rhoa} that 
	\begin{align}
	\rho_a(T_0) \propto \frac{1}{\Tosc^3}\left( \frac{\Tosc}{\Trh}   \right)^{\frac{3\alpha - 3}{\alpha}} =\Trh^{\frac{3-3\alpha }{\alpha}} \Tosc^{-\frac{3}{\alpha}} \propto  
	\Trh^{\frac{4-n}{n} } m_a^{-\frac{n+2}{n}} \,,
	\end{align}
	where we have utilized Eq.~\eqref{eq:Tosc} to rewrite $\Tosc$ with $m_a$ and $\Trh$ in the last step.  We also notice that for $n=4$, the relic abundance is independent from the reheating $\Trh$, being  the same as the first line of Eq.~\eqref{eq:Omega_RD}. 
	\begin{figure}[ht]
		\def\sepf{0.4}
		\centering
		\includegraphics[scale=\sepf]{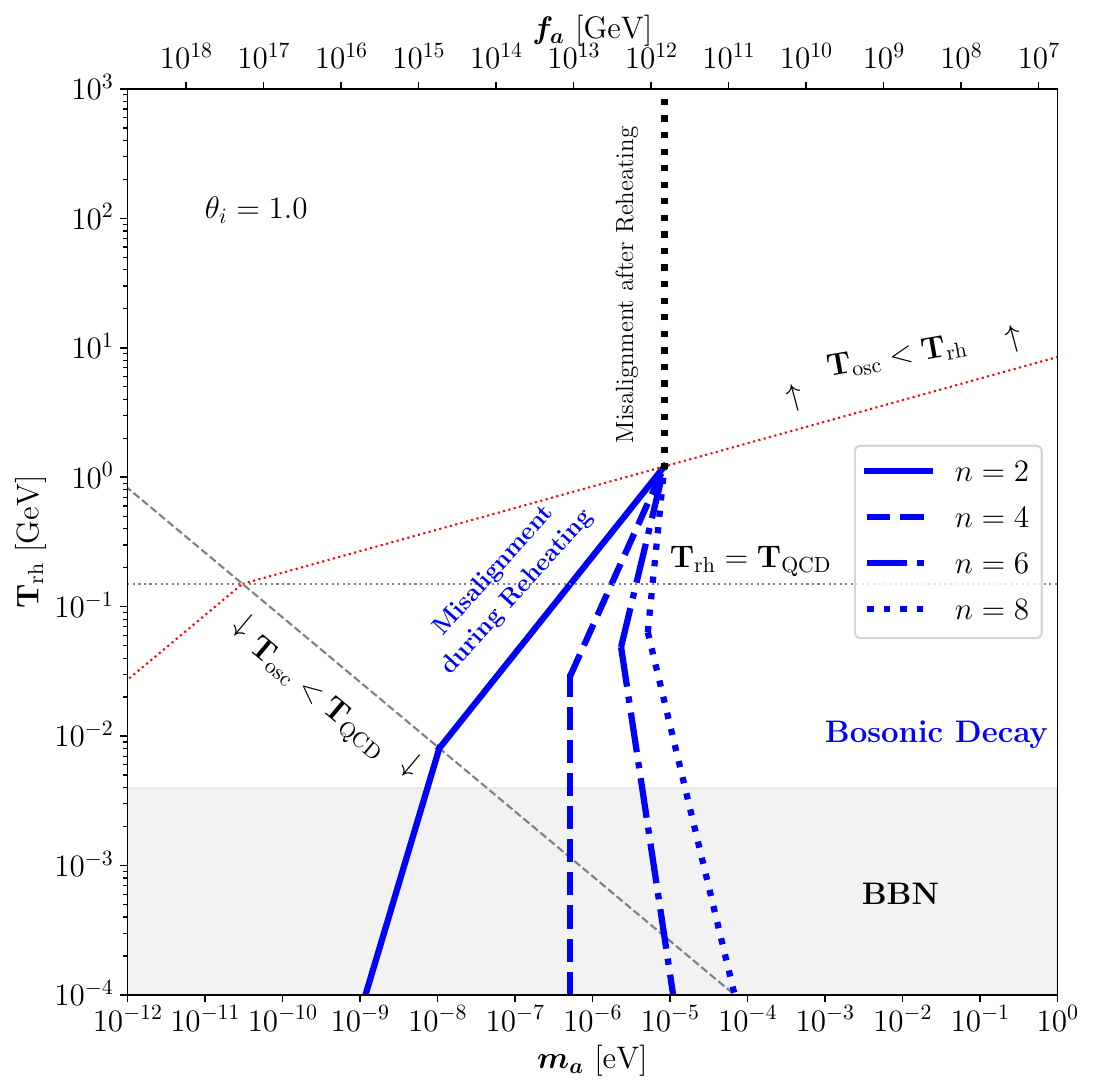}
		\includegraphics[scale=\sepf]{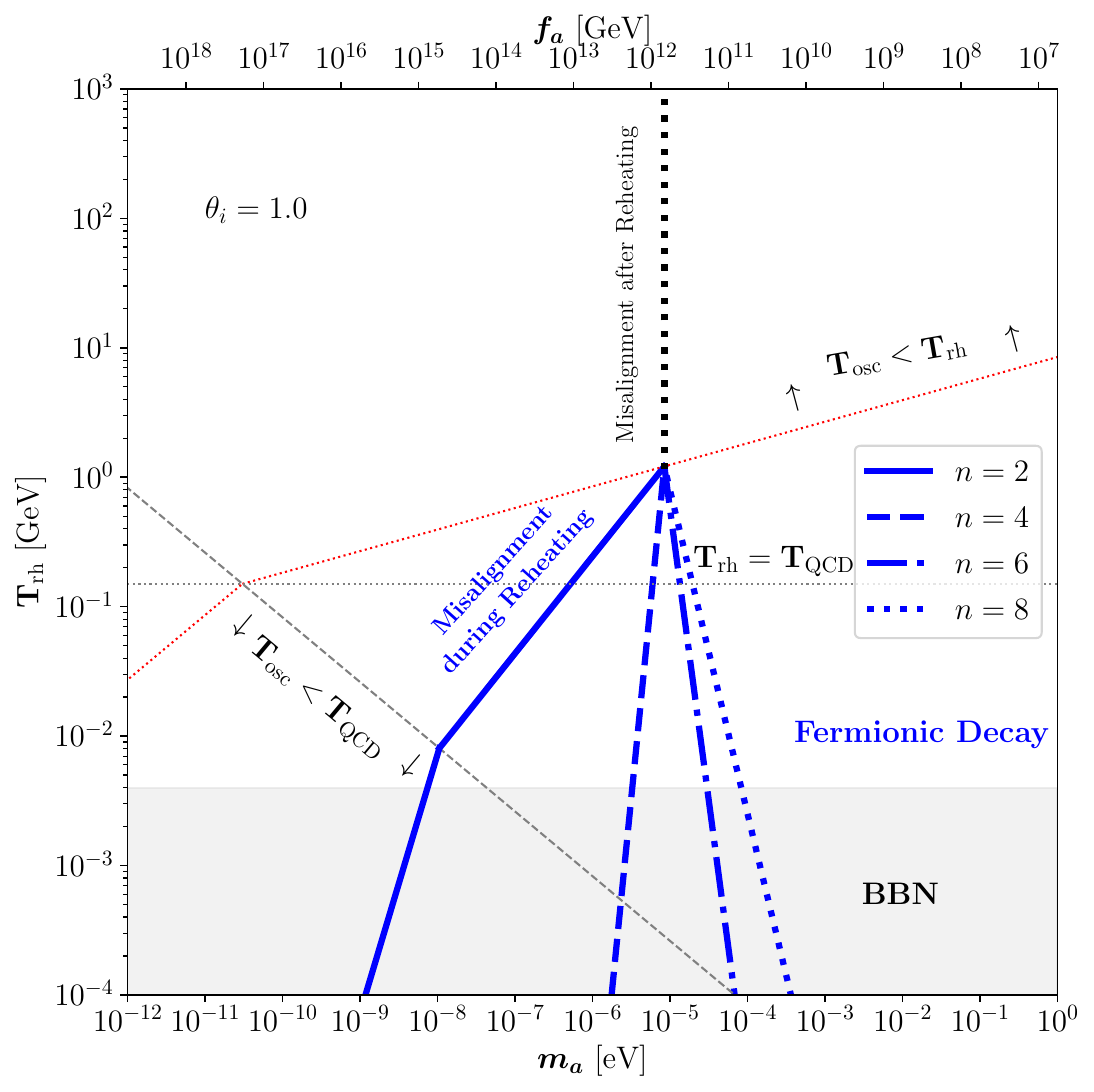}
		\includegraphics[scale=\sepf]{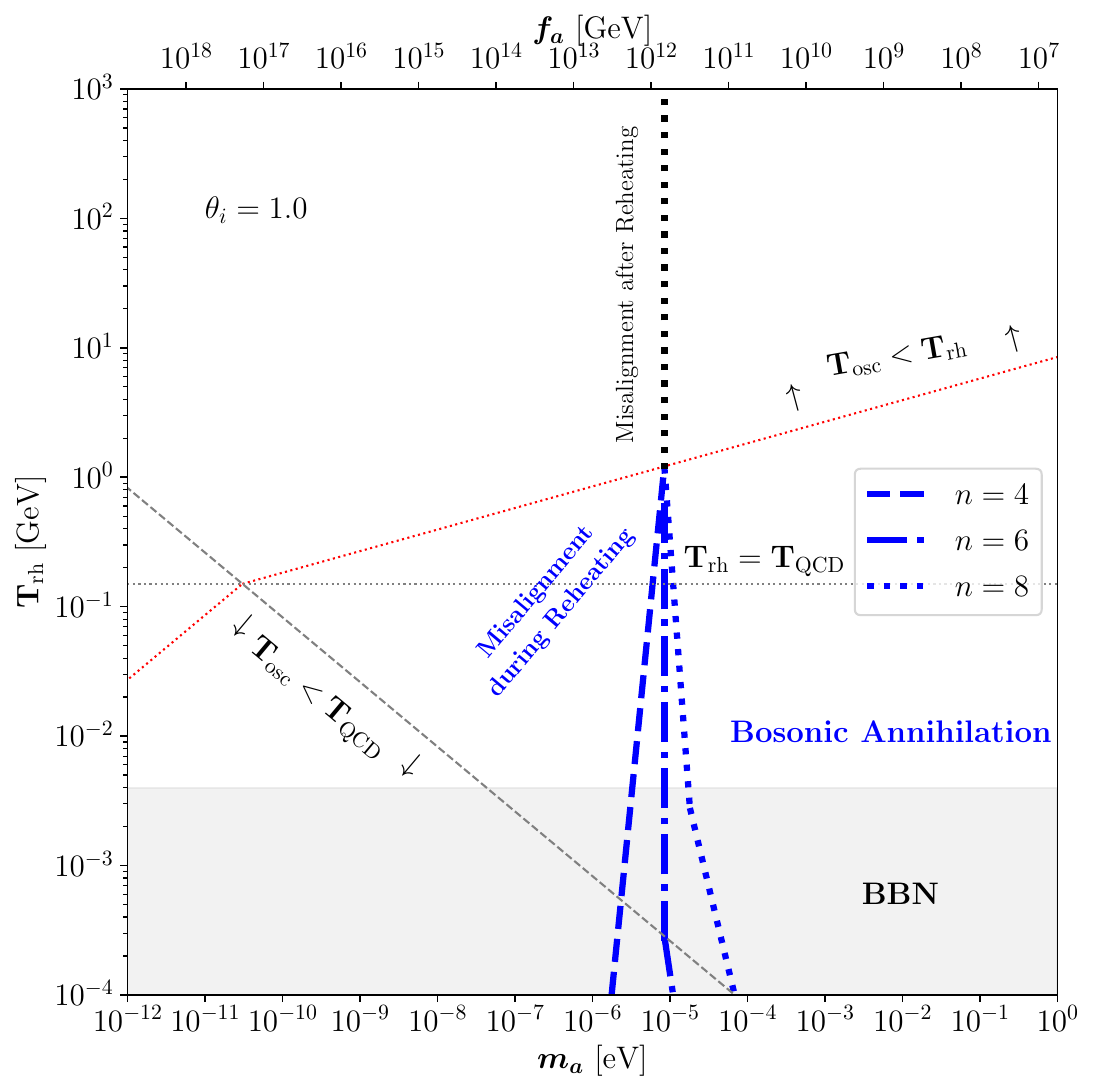}
		\caption{Blue lines correspond to $\Trh$ as function of $m_a$ (or $f_a$)  generating the observed axion DM relic density for bosonic (upper left), fermionic (upper right) inflaton  decay and bosonic inflaton annihilation (lower panel) with different $n$ and $\theta_i=1$. }
		\label{fig:case2}
	\end{figure} 

	In Fig.~\ref{fig:case2}, we illustrate the parameter space $(\Trh, m_a)$ or $(\Trh, f_a)$ that gives rise to the observed DM relic abundance. The blue lines, depicted as solid, dashed, dash-dotted, and dotted, correspond to different values of $n$, specifically $n=2, 4,\, 6~\text{and}~8$, respectively. The red dotted line corresponds to $\Trh = \Tosc$, which divides the parameter space into regions of $\Trh > \Tosc$ and $\Trh < \Tosc$, above and below it, respectively. The vertical black dotted line represents the scenario with $\Trh > \Tosc$, where oscillations occur \textit{after} reheating, specifically during the radiation epoch, with $m_a \simeq 8.6 \cdot 10^{-6}\text{eV}$. The horizontal gray dotted line denotes $\Trh = \Tqcd$. The gray dashed line symbolizes $\Tosc = \Tqcd$, marking the boundary between regions of $\Tosc < \Tqcd$ and $\Tosc > \Tqcd$. Note that $\Tosc$ depends on both $n$ and $\alpha$ (or the type of inflaton-matter couplings). For a fixed $n$, the proximity of $\Tosc$ to $\Tqcd$ leads to a change in slope for the blue lines, an effect attributed to the temperature-dependent nature of the axion mass. In the plot, we only depict $\Tosc = \Tqcd$ (gray dashed line) with $n=2$ for bosonic decay (upper left panel) and fermionic decay (upper right panel), and with $n=6$ for bosonic annihilation (lower panel). For the fermionic decay with $n>2$ and bosonic annihilation with $n=4$, the blue lines change slopes at regimes with much smaller $\Trh$, which are not visible in the figure. Lastly, the shaded gray band characterizes $\Trh \lesssim 4~\text{MeV}$, a range contradictory to the requirements of Big Bang nucleosynthesis (BBN)~\cite{Kawasaki:2000en, Hannestad:2004px}, and therefore disfavored.
	
	In the scenario where $n=2$, we observe that the parameter space remains identical for both bosonic and fermionic decay modes, as explained before. In this context, the axion mass can reach remarkably small values, approximately $m_a \simeq 7.5\cdot10^{-9}\,\text{eV}$, assuming $\Trh \simeq 4\,\text{MeV}$ and $\theta_i = 1.0$. By further reducing $\theta_i$, the lower limit of $m_a$ can be pushed even lower. For example, with $\theta_i = 0.5$, the axion mass can be as small as $m_a \simeq 3.7\cdot10^{-9}\,\text{eV}$ while maintaining $\Trh \simeq 4\,\text{MeV}$. Conversely, if we explore larger misalignment angles, such as $\theta_i = \pi/\sqrt{3}$, and higher reheating temperatures, around $\Trh \simeq 1\,\text{GeV}$, we find that the axion mass increases, reaching approximately $m_a \simeq 2.4\cdot10^{-5}\,\text{eV}$. These results for $n=2$ align with the early matter case discussed in Ref.~\cite{Arias:2021rer}. 
	
	As we examine larger values of $n$ (while keeping $\Trh$ constant), an intriguing trend emerges. The dilution factor, as presented in Eq.~\eqref{eq:dilution_S}, tends to grow, leading to an increased demand for a larger axion mass to satisfy the condition that the axion density at present is fixed. This elucidates the ordering of the blue curves (from left to right) within the parameter space. Although we are focusing on the regime with $n\leq 8$, it is still interesting to note that in the limit $n\to \infty$, axion mass can be as large as $\mathcal{O}(10^{-2})~\text{eV}$ (cf. Eq.~\eqref{eq:Omega_a}).  This upper bound on axion mass is in agreement with scenarios where oscillations occur during the kination epoch \cite{Arias:2021rer}.
	
	With the increase of the reheating temperature $\Trh$, 
	we notice that the blue lines tend to merge to the black dotted line. The physical reason is that for larger $\Trh$, misalignment tends to occur {\it after} reheating so that the effect of details of reheating dynamics do not play a role. Using Eq.~\eqref{eq:Trhbound2}, and $m_a \simeq 8.6\cdot 10^{-6}~\text{eV}$, we find  an upper bound on reheating temperature:
	\begin{align}
	\Trh \lesssim  1~\text{GeV}\,,
	\end{align} which guarantees that misalignment happens during reheating.
	
	An  intriguing feature of Fig.~\ref{fig:case2} is that the distinctions among different reheating scenarios as well as their effects on the extended parameter space become pronounced for $n>2$, which contribute to the distinct shapes of the curves as depicted in the three figures. As a result, we have different ranges of axion mass for a given reheating temperature $\Trh \lesssim 1~\text{GeV}$ in the  different reheating scenarios.

	\begin{table}[ht]
		\begin{tabular}{|l||*{3}{c|}}\hline
			\backslashbox[34mm]{$n$}{\rotatebox{-20}{$m_a/\text{eV}$}}{\rotatebox{0}{scenario}}
			&\makebox[3em]{Ferminonic Decay}&\makebox[3em]{Bosonic Decay}&\makebox[3em]{Bosonic Annihilation}
			% &\makebox[3em]{6/3}
			\\\hline \hline
			$n=2$ & $\left[3.7\cdot 10^{-9}, 2.4 \cdot 10^{-5} \right]  $ &$\left[3.7\cdot 10^{-9}, 2.4 \cdot 10^{-5} \right] $&\\\hline \hline
			%%%%%%%%%%%%%%%%%%%%%%%%%%%%%%%%%%%%
			$n=4$  &$\left[5.1\cdot 10^{-7},  2.4 \cdot 10^{-5}  \right] $&$\left[2.1 \cdot 10^{-7}, 2.4 \cdot 10^{-5}  \right] $&$\left[5.1\cdot 10^{-7}, 2.4 \cdot 10^{-5} \right] $\\\hline \hline
			%%%%%%%%%%%%%%%%%%%%%%%%%%%%%%%%%%%%
			$n=6$  &$\left[2.6 \cdot 10^{-6}, 8.7 \cdot 10^{-5} \right] $&$\left[1.6 \cdot 10^{-6}, 2.4 \cdot 10^{-5} \right] $&$  \left[2.6 \cdot 10^{-6}, 2.4 \cdot 10^{-5}\right] $\\\hline \hline
			%%%%%%%%%%%%%%%%%%%%%%%%%%%%%%%%%%%%
			$n=8$  &$\left[2.6 \cdot 10^{-6},  2.5 \cdot 10^{-4}\right] $&$\left[2.6 \cdot 10^{-6}, 4.1 \cdot 10^{-5} \right] $& $\left[2.6 \cdot 10^{-6} , 4.9 \cdot 10^{-5} \right] $ \\\hline \hline
		\end{tabular} 
		\caption{Summary of parameter space of $m_a$ for QCD axion DM 
			in a different reheating scenario with $\theta_i \in \left[0.5, \pi/\sqrt{3} \right]$ and $4~\text{MeV}\lesssim \Trh  \lesssim 1~\text{GeV}$.}
		\label{Tab:ma}
	\end{table}

	So far, we have only focused on $\theta_i=1.0$. In Table \ref{Tab:ma}, we explore the parameter space concerning the axion mass $m_a$ for various reheating scenarios and specific values of $n$ by allowing wider range of initial misalignment angle: $\theta_i \in \left[0.5, \pi/\sqrt{3} \right]$.  In the bosonic annihilation scenario, we leave it empty for $n=2$ since it is not possible to have successful reheating with  inflaton annihilation alone. Again, one can see that the parameter space can vary for different reheating scenarios. In particular, the bosonic annihilation scenarios  do {\it not} allow axion mass in the regime: $3.7\cdot 10^{-9 }~\text{eV} \lesssim m_a \lesssim 5.1\cdot 10^{-7} ~\text{eV}$, which is nevertheless supported in the decay scenarios. It is also interesting to note that only fermionic decay scenarios can give rise to $ 4.9 \cdot 10^{-5 }~\text{eV} \lesssim m_a \lesssim 2.5 \cdot 10^{-4} ~\text{eV}$. These  distinctions in the parameter space have very  important experimental implications, as will be explored in Sec.~\ref{sec:exp}.
	
	%%%%%%%%%%%%%%%%%%%%%%%%%%%%%%%%%%
	\subsection{ALP}
	%%%%%%%%%%%%%%%%%%%%%%%%%%%%%%%%%%
	So far, we have only focused on the QCD axions. In this section, we shift our focus to the axion-like particles (ALPs). When oscillations happen {\it during} reheating, the oscillation temperature is the same as  the expression given in Eq.~\eqref{eq:Tosc_case2}.  The allowed  reheating temperature in this case reads 
	\begin{align}\label{eq:ALPTrh_bound}
	4~\text{MeV}\leq \Trh \leq \Tosc \simeq  \left(\frac{1}{\pi} \sqrt{\frac{10}{\gs}} m_a\,M_P\right)^{\frac{1}{2}}\,,
	\end{align} 
	which leads to a lower bound on ALP mass:
	\begin{align}\label{eq:m_ALP_bound}
	m_a \gtrsim 2.2\cdot 10^{-14}~\text{eV}\,.
	\end{align}
	Here we assume that ALP mass to be a constant in time, and the ALP energy density  is then given by
	\begin{align} \label{eq:_rhoALP2}
	\rho_a(T_0) = \rho_a(\Tosc)  \frac{s(T_0)}{s(\Tosc)}  \frac{S(\Tosc)}{S(\Trh)} \simeq \frac12 m_a^2\, f_a^2\, \theta_i^2\, \frac{s(T_0)}{s(\Tosc)}  \frac{S(\Tosc)}{S(\Trh)}\,,
	\end{align} 
	where the entropy dilution factor $S(\Tosc)/S(\Trh)$ can be computed using Eq.~\eqref{eq:dilution_S}. This leads to:
	\begin{align}\label{eq:_rhoALP3}
	\Omega_ah^2 &\simeq \left(\frac{\theta_i}{1.0}\right)^2  \left(\frac{f_a}{1.3 \cdot 10^{13} \cdot \text{exp}(8.1/n)~\text{GeV}}\right)^2\, \left(\frac{\Trh}{4~\text{MeV}}\right)^{\frac{4-n}{n}}  \left(\frac{m_a}{7.4\cdot 10^{-11}~\text{eV}}\right)^{\frac{n-2}{n}} \,,
	\end{align}
	or in specific value of $n$:
	\begin{align}\label{eq:_rhoALP3_n}
	\Omega_ah^2 \simeq
	\left(\frac{\theta_i}{1.0}\right)^2 
	\begin{dcases}
	\left(\frac{f_a}{7.6 \cdot 10^{14}~\text{GeV}}\right)^2  \left( \frac{\Trh}{4~\text{MeV}}\right)^{1}  \left( \frac{m_a}{7.4\cdot 10^{-11} ~ ~\text{eV} }\right)^{0}
	& \text{$n=2$,} \\
	\left(\frac{f_a}{1.0 \cdot 10^{14}~\text{GeV}}\right)^2 \left( \frac{\Trh}{4 ~\text{MeV}}\right)^{0}  \left( \frac{m_a}{7.4\cdot 10^{-11} ~\text{eV} }\right)^{\frac{1}{2}}& \text{$n=4$,} \\
	\left(\frac{f_a}{5.1 \cdot 10^{13}~\text{GeV}}\right)^2  \left( \frac{\Trh}{4~\text{MeV}}\right)^{-\frac{1}{3}}  \left( \frac{m_a}{7.4\cdot 10^{-11} ~\text{eV} }\right)^{\frac{2}{3}} & \text{$n=6,$} \\
	\left(\frac{f_a}{3.6 \cdot 10^{13}~\text{GeV}}\right)^2 \left( \frac{\Trh}{4~\text{MeV}}\right)^{-\frac{1}{2}}  \left( \frac{m_a}{7.4\cdot 10^{-11}~\text{eV} }\right)^{\frac{3}{4}} & \text{$n=8.$} \\
	\end{dcases}
	\end{align}

	As a means of cross checking and to ensure the consistency of our analysis, we note that Eq.~\eqref{eq:_rhoALP3} and Eq.~\eqref{eq:_rhoALP3_n} reproduce Eq.~\eqref{eq:Omega_a} and Eq.~\eqref{eq:Omega_a_n} once replacing $f_a$ to $m_a$ by using Eq.~\eqref{eq:ma1}.  Similar to the case with $\Tosc < \Tqcd$ for  QCD axion, we find that the parameter space remains unaffected by the type of inflaton-matter couplings under consideration.

	Our results in this section become particularly interesting when examining different values of $n$ within the context of ALPs. Starting with the case where $n=2$, our analysis aligns with the results presented in Ref.~\cite{Blinov:2019rhb}, where the authors study ALPs misalignment during early matter domination. We confirm that $ \Omega_ah^2$ remains independent of $m_a$, thereby resulting in the relationship $f_a \propto \Trh^{-1/2}$ as a prerequisite for achieving the observed DM relic abundance. Another scenario, outlined in Ref.~\cite{Blinov:2019rhb}, corresponds to the limit of $n\to \infty$, where we find that $f_a \propto m_a^{-1}$ holds true. Our results for $n=2$ and $n \to \infty$ thus stand as a validation of the earlier work. 
	
	Being complementary to Ref.~\cite{Blinov:2019rhb}, we explore additional values of $n$, revealing the trends as follows.
	\begin{itemize}
		\item For $n=4$, $\Omega_ah^2$ becomes independent of $\Trh$, mirroring the scenario where ALP oscillates during the radiation epoch (cf. Eq.~\eqref{eq:Omega_a_case2}). Consequently, to account for the observed DM relic abundance, the requirement shifts to $f_a \propto m_a^{-1/4}$.
		\item When considering $n>4$, we delve into a domain where $f_a$ can be significantly smaller than what would be necessary under radiation-like oscillation conditions. For instance, for $n=6$, we ascertain that $f_a \propto m_a^{-1/3}$, while for $n=8$, the relationship shifts to $f_a \propto m_a^{-3/8}$.
	\end{itemize}
	This intriguing range of $n$ values leads to interesting experimental implications that will be explored in the forthcoming section.
	
	%%%%%%%%%%%%%%%%%%%%%%%%%%%%%%%%%%%
	\section{Experimental Constraints}\label{sec:exp}
	%%%%%%%%%%%%%%%%%%%%%%%%%%%%%%%%%%%
	In previous sections, we have worked out the parameter space for QCD axion and ALP misalignment {\it during} reheating. We have shown that the parameter space can be enlarged and show distinct features for different reheating scenarios. In this section, we are devoted to investigating the experimental constraints and implications.
	
	As a benchmark scenario, we consider axions and ALPs  couple to two photons via an  effective dimension-5 operator ~\cite{Marsh:2015xka,  Graham:2015ouw, DiLuzio:2020wdo, Sikivie:2020zpn}:
	\begin{align}
	\mathcal L_{a\gamma} = -\frac14\, g_{a\gamma}\, a\, F_{\mu\nu}\, \tilde F^{\mu\nu} = g_{a\gamma}\, a\, \vec E \cdot \vec B\,,
	\end{align}
	where $g_{a\gamma}$ corresponds to the coupling constant. It is related to the decay constant as~\cite{Graham:2015ouw}
	\begin{align}\label{eq:g_agamma}
	g_{a\gamma} 
	\simeq 10^{-13}~\text{GeV}^{-1} \left(\frac{10^{10}~\text{GeV}}{f_a}\right).
	\end{align}
	
	In the landscape of experimental constraints, numerous developments have emerged to shed light on the axion and ALP parameter space~\cite{Graham:2015ouw, Irastorza:2018dyq, Adams:2022pbo}. Within this context, our primary focus centers on specific experimental endeavors, particularly Haloscope and Telescope experiments, which are relevant to the parameter space of interest in this article. Here we briefly mention several relevant experiments, including some ongoing Haloscope experiments, e.g. ADMX \cite{ADMX:2018gho, ADMX:2019uok, Crisosto:2019fcj, ADMX:2021nhd, ADMX:2021mio} and  CAPP \cite{Lee:2020cfj, Jeong:2020cwz, CAPP:2020utb, Lee:2022mnc, Kim:2022hmg} and Telescope experiments e.g. MUSE \cite{Todarello:2023hdk}. The sensitivity regimes are depicted in dark gray color. We also consider some future projection limits in gray dashed lines from from several Haloscope experiments aimed for different mass regimes. These include DM-Radio for $m_a$ in $20-800 \, \text{neV}$ \cite{DMRadio:2022pkf}, FLASH for $m_a$ in  $0.2 - 1 \, \mu \text{eV}$  \cite{Alesini:2019nzq}, Baby-IAIXO for $m_a$ in  $1-2\, \mu \text{eV}$  \cite{Diaz-Morcillo:2021psa}, ADMX for $m_a$ in  $1.9-3.7\, \mu \text{eV}$ \cite{Stern:2016bbw}, QUAX for $m_a$ in  $35-45 \, \mu \text{eV}$,  DALI for $m_a$ in    $20-250\, \mu \text{eV}$ \cite{DeMiguel:2023nmz}, ALPHA for $m_a$ in    $35-  400 \, \mu \text{eV}$ \cite{Lawson:2019brd},  MADMAX for $m_a$ in    $40-  400 \, \mu \text{eV}$ \cite{Beurthey:2020yuq},  ORGAN  for $m_a$ in  $60 - 210 \, \mu \text{eV}$ \cite{McAllister:2017lkb}, CADEx  for $m_a$ in   $330- 460\,  \mu \text{eV}$ \cite{Aja:2022csb}, BRASS for $m_a$ in    $10  -104 \, \text{meV}$, BREAD for $m_a$ in    $10^{-3}  -1\, \text{eV}$ \cite{BREAD:2021tpx}, LAMPOST for $m_a$ in   $0.1  -10\, \text{eV}$  \cite{Baryakhtar:2018doz}.  These constraints and projection curves have been generated by using the  AxionLimits code in \cite{AxionLimits}.
	
	\subsection{Results}
	\begin{figure}[ht]
		\def\sepf{0.39}
		\centering
		\includegraphics[scale=\sepf]{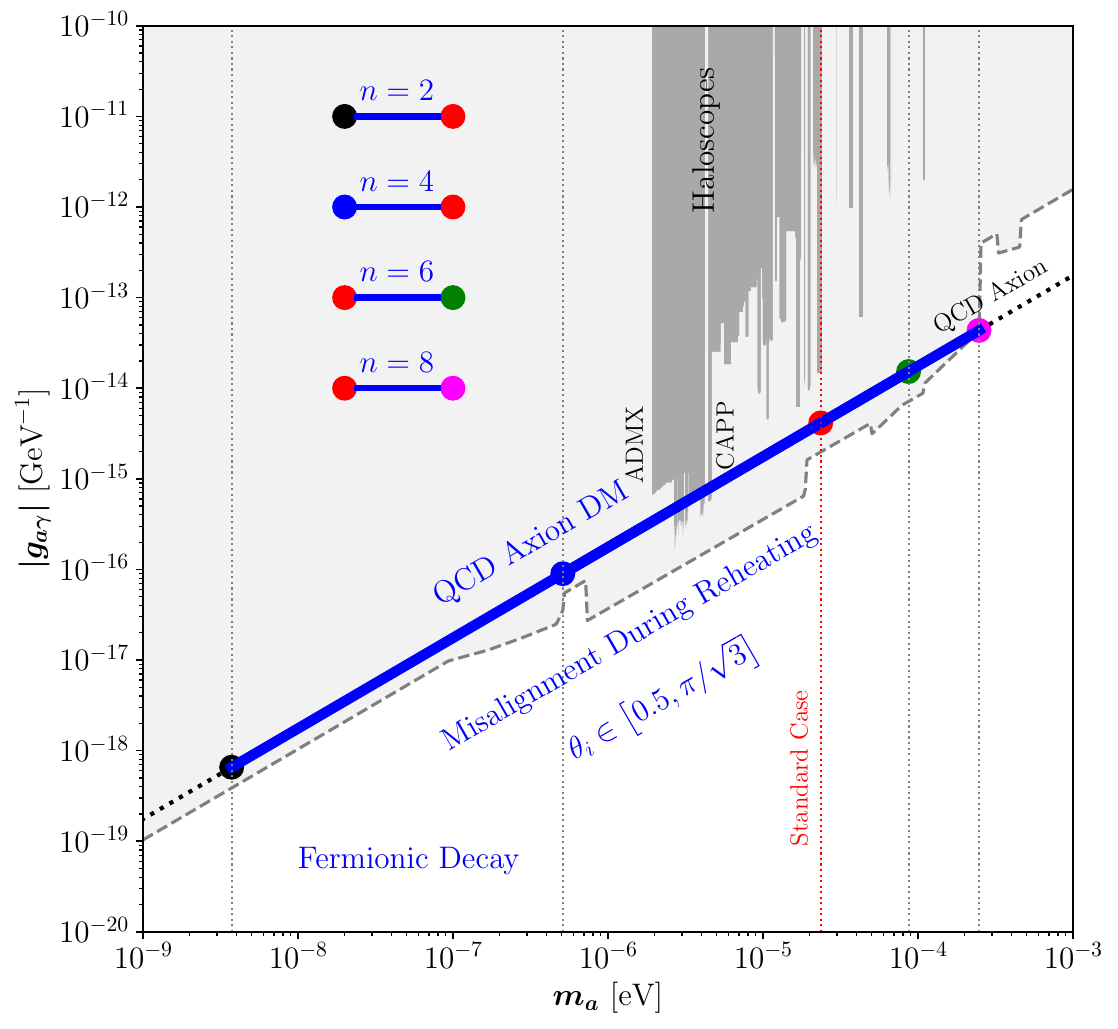}
		\includegraphics[scale=\sepf]{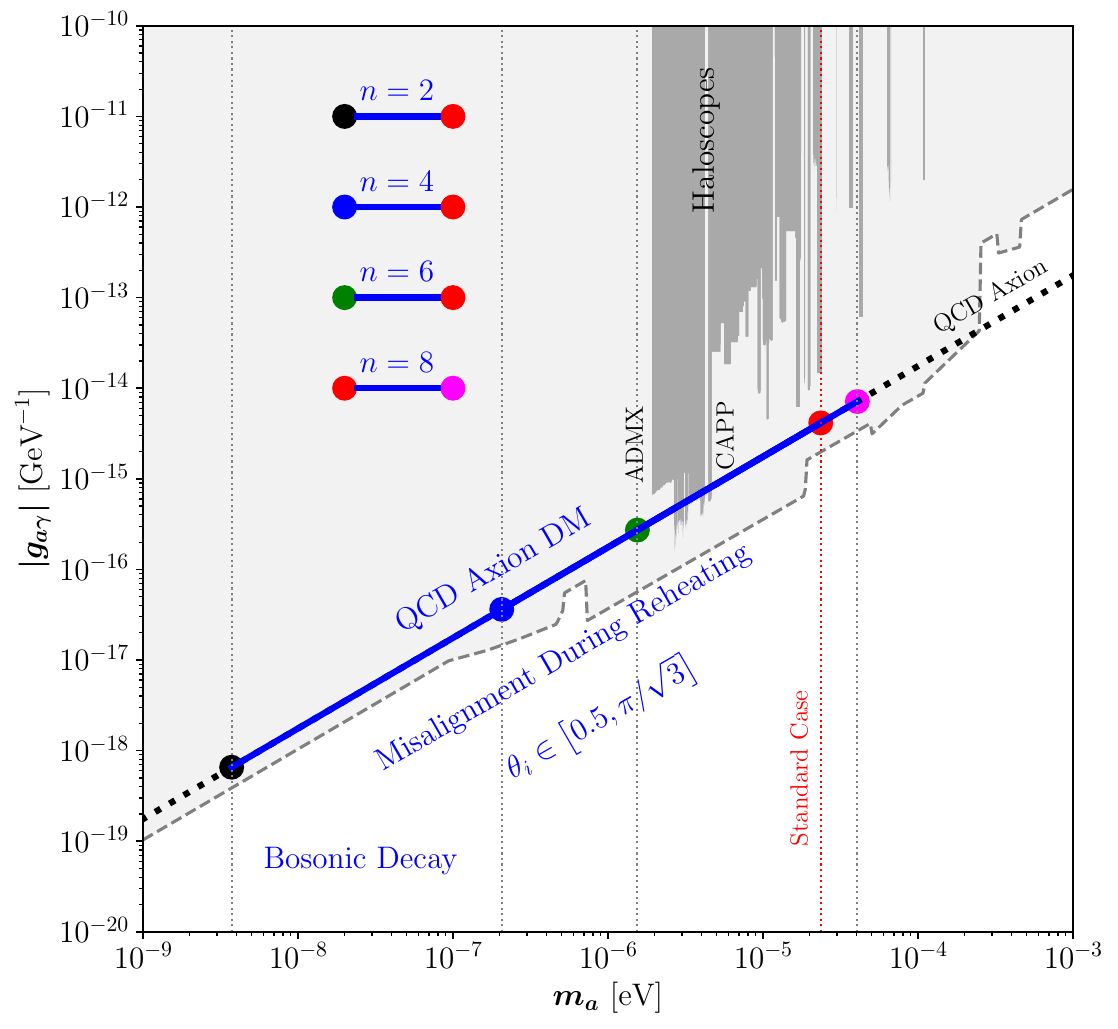}
		\includegraphics[scale=\sepf]{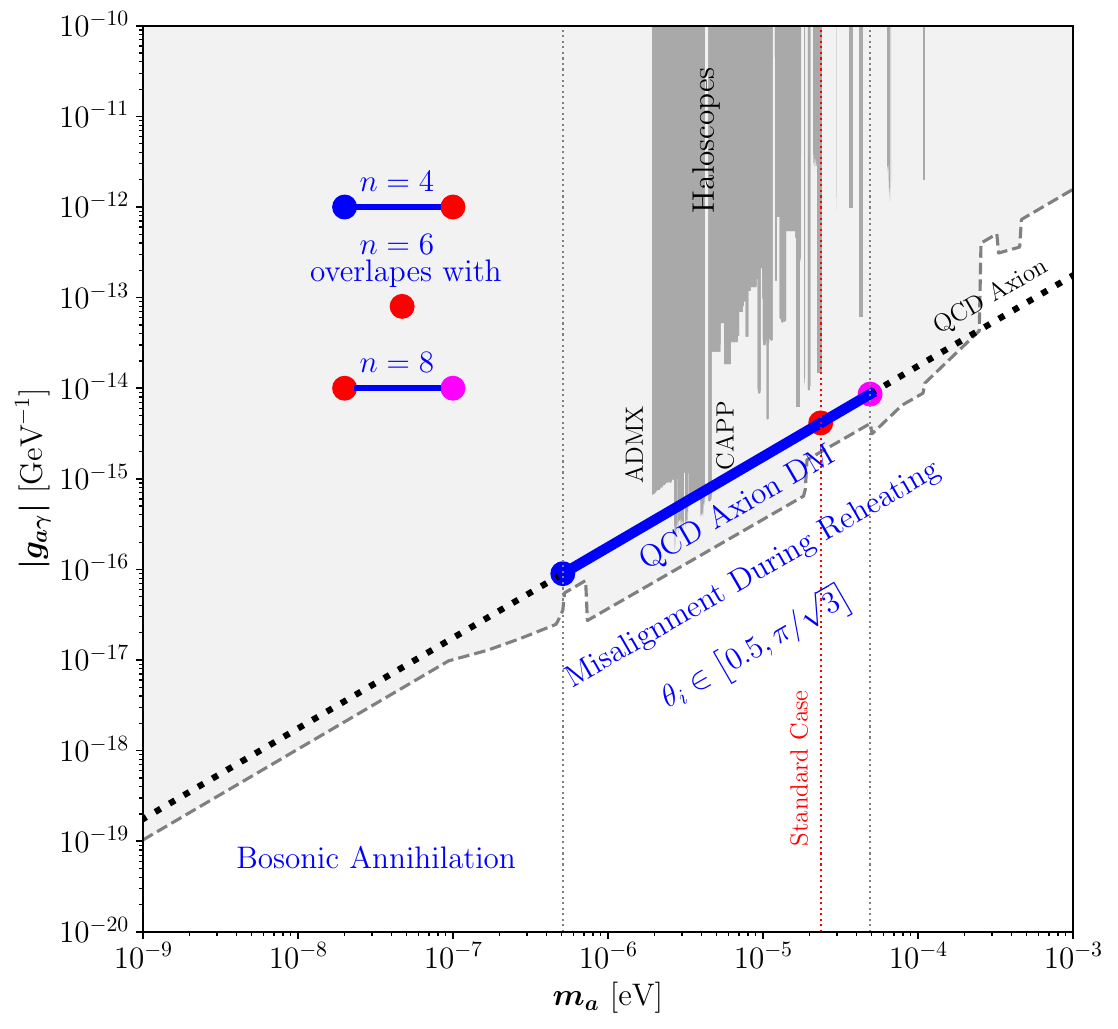}
		\caption{The blue lines correspond to the parameter space with $g_{a\gamma}$ as function of $m_a$ to yield correct DM relic abundance for QCD axion by assuming misalignment occurs {\it during} reheating. The red dot corresponds to the standard case where oscillation occurs in radiation epoch {\it after} reheating with $\theta_i = \pi/\sqrt{3}$. The segments between red and black, red and blue, red and green, red and magenta dots correspond to the parameter space with $n=2$, $n=4$, $n=6$, and $n=8$, respectively. For bosonic annihilation, $n=6$ overlaps with the standard case (red dot).   The constraints (dark grey region) and future projection limits (grey dashed lines)  are generated by using the  AxionLimits code \cite{AxionLimits}.}
		\label{fig:QCDgag}
	\end{figure} 
	
	In Fig.~\ref{fig:QCDgag}, we present the parameter space $(g_{a\gamma}, m_a)$ that leads to the correct relic abundance. The black dotted line represents the prediction for the QCD axion using Eq.~\eqref{eq:ma1} and Eq.~\eqref{eq:g_agamma}. It is important to note that not all points along this line fulfill the condition for axions to produce the correct relic abundance.  For the scenario where oscillations occur during the radiation epoch \textit{after} reheating, we find specific values that match this requirement: $m_a \simeq 2.4 \cdot 10^{-5}~\text{eV}$ and $g_{a\gamma} \simeq 4.1\cdot 10^{-15}~\text{GeV}^{-1}$ with $\theta_i = \pi/\sqrt{3}$. This specific point is illustrated as the red dot on the plot. It is worth mentioning that the parameter space can be expanded further if the misalignment takes place {\it during} the reheating phase, as thoroughly investigated in the previous section.

	The upper left, upper right, and lower panels of the figure correspond to distinct reheating scenarios achieved via inflaton fermionic decay, inflaton bosonic decay, and bosonic annihilation, respectively. In each panel, the blue lines delineate the parameter space where the QCD axion can account for dark matter, specifically when oscillations occur \textit{during} the reheating phase. The segments between red dot and other dots of various colors, namely black, blue, green, and magenta, delineate the regions where the conditions for the observed relic abundance are satisfied for different values of $n$ ($n=2$, $n=4$, $n=6$, and $n=8$).   The gray vertical dotted lines in all panels correspond to the lower and upper limits on $m_a$ as outlined in Table \ref{Tab:ma}.

	We have considered the initial misalignment angle as $\theta_i \in \left[0.5, \pi/\sqrt{3}\right]$ in all panels. 
	It is interesting to notice that a portion of the extended parameter space are already  constrained  by the Haloscope experiments, for example ADMX \cite{ADMX:2018gho, ADMX:2019uok, Crisosto:2019fcj, ADMX:2021nhd, ADMX:2021mio} and  CAPP \cite{Lee:2020cfj, Jeong:2020cwz, CAPP:2020utb, Lee:2022mnc, Kim:2022hmg}.  Depending on the underlying reheating scenarios, the conclusions differ as will be explained in the following.
	
	\begin{itemize}
		\item Fermionic Decay Scenario.
		
		For the reheating scenario through inflaton fermionic decay, we observe that the parameter space corresponding to $n=2$ and $n=4$ is susceptible to sensitivity by ADMX and CAPP experiments. Notably, regions within the axion mass range of $2\cdot10^{-6}~\text{eV}$ to $5 \cdot10^{-6}~\text{eV}$ are already ruled out by these experiments, as depicted in the upper left panel of  Fig.~\ref{fig:QCDgag}. However, for $n>4$, the current Haloscope experiments are less likely to probe the parameter space associated with the reheating scenario via fermionic decay.
		
		\item  Bosonic Decay Scenario. 
		
		Within the context of the bosonic decay scenario, the results are different compared to previous fermionic decay scenarios. Specifically, the current ADMX and CAPP experiments are  sensitive to the scenarios with $n\leq 6$. 
		
		\item  Bosonic Annihilation Scenario.
		
		In the context of the bosonic annihilation scenario with $n=6$, the parameter space overlaps with the case of oscillations during the radiation epoch.
		The parameter space for $n=4$ is identical to that in the fermionic decay scenario and is slightly larger than that in the bosonic decay scenario for $n=8$, which is also shown in Table \ref{Tab:ma}. It is evident that only the scenario with $n=4$ is constrained with the current ADMX and CAPP experiments.
	\end{itemize}
	\begin{figure}[ht]
		\def\sepf{0.8}
		\centering
		\includegraphics[scale=\sepf]{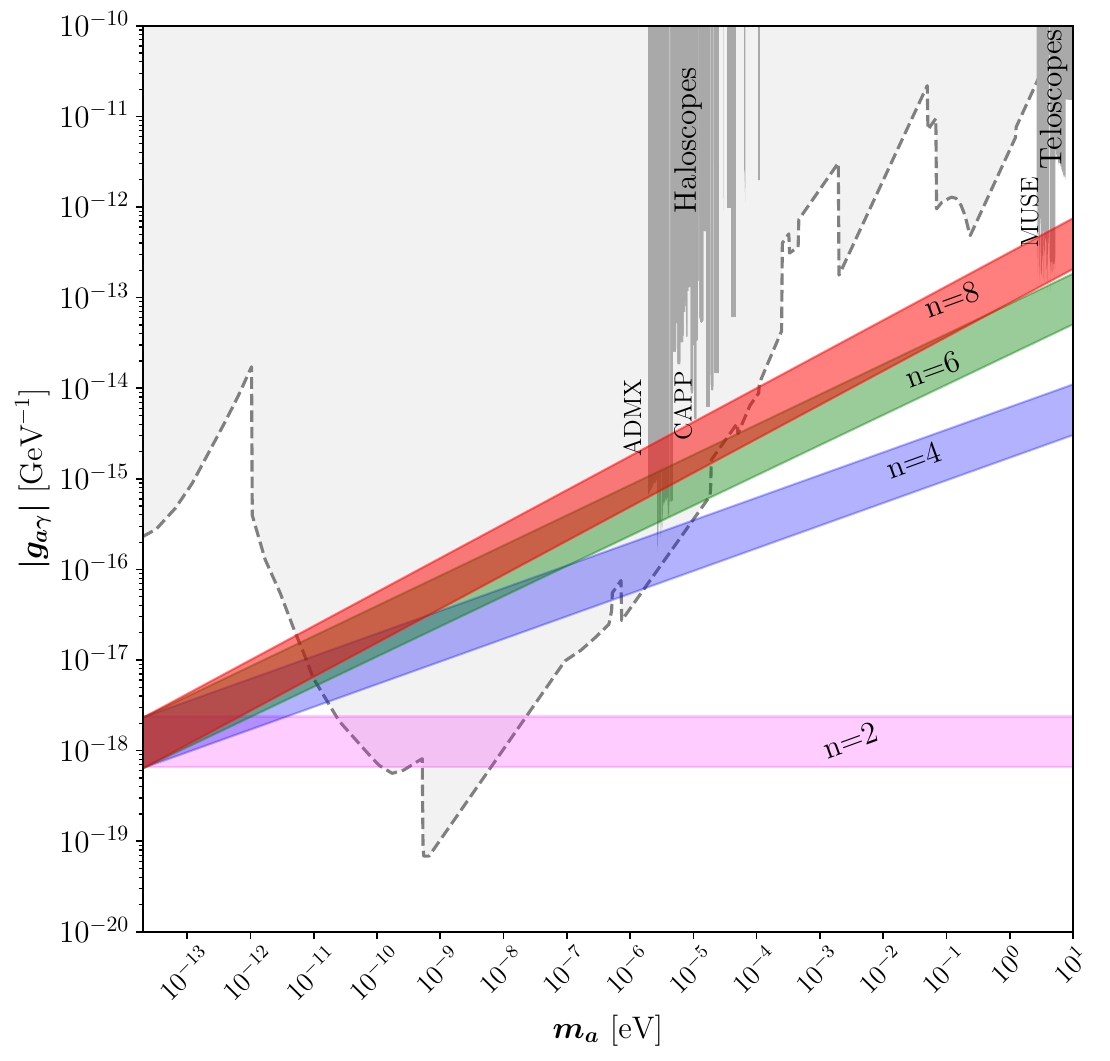}
		\caption{Colored bands correspond to the parameter space for $g_{a\gamma}$ as function of $m_a$ to yield correct ALP DM relic abundance by assuming misalignment occurs {\it during} reheating with different values of $n$. We have considered $\theta_i \in \left[0.5, \pi/\sqrt{3}\right]$ and $\Trh = 4~\text{MeV}$.  The scenario with $n=4$ is identical to the standard case where oscillations occur during the radiation epoch.
			The current constraints (dark grey regimes) and future projection limits (grey dashed lines)  are generated by using the  AxionLimits code \cite{AxionLimits}.
		}
		\label{fig:gag}
	\end{figure} 

	In Fig.~\ref{fig:gag}, we illustrate the relationship between $(g_{a\gamma}, m_a)$ that results in the correct relic abundance for ALPs, with a focus on its dependence on the parameter $n$. We have fixed $\theta_i \in \left[0.5, \pi/\sqrt{3}\right]$ and $\Trh = 4~\text{MeV}$, and considered the ALP mass bound presented in Eq.~\eqref{eq:m_ALP_bound}, as we are mainly focusing on misalignment during reheating.  
	
	The magenta, blue, green and red shaded bands correspond to different values of $n$, specifically $n=2$, $n=4$, $n=6$, and $n=8$, respectively. For fixed $m_a$ and reheating temperature $\Trh$, with increasing values of $n$, a smaller decay constant $f_a$ is required (cf. Eq.~\eqref{eq:_rhoALP3_n}), which in turn leads to a larger value of the axion-photon coupling constant $g_{a\gamma}$. Consequently, the parameter space corresponding to higher values of $n$ becomes more accessible for probing. For larger reheating temperature, the red band with $n=8$ and green band with $n=6$ move downwards, while the magenta band with $n=2$ moves upwards. Note that the case with $n=4$ is identical to the scenario where oscillations occur during the radiation epoch, being independent of $\Trh$.

	An intriguing observation emerges when considering the impact of ongoing experiments. For $n=8$, current Haloscope experiments, such as ADMX and CAPP, have already constrained the parameter space within the axion mass range of $2\cdot10^{-6}~\text{eV}$ to $5 \cdot10^{-6}~\text{eV}$. Additionally, Telescope experiments, like MUSE \cite{Todarello:2023hdk}, are capable of constraining the parameter space with $2.7~\text{eV}\lesssim m_a \lesssim 5.3~\text{eV}$ for this case. As $n$ decreases, for instance in the case of $n=6$, the parameter space moves beyond the sensitivity of MUSE but can still be probed through ADMX and CAPP experiments. For $n=4$ and $n=2$, the parameter space extends beyond the reach of current ADMX and CAPP experiments, resulting in a different conclusion compared to the scenario of QCD axions.  
	
	\subsection{Probing Reheating via Axion  Experiments}\label{sec:reheating_constraint}
	Given the distinct  behaviors of the parameter space and the resultant constraints arising from different reheating scenarios, it is possible that future axion experiments hold considerable potential for constraining the reheating process. Here, we offer insights into how these experiments could play a role in constraining the reheating dynamics.
	
	For the QCD axion, it is evident from Fig.~\ref{fig:QCDgag} that future Haloscope experiments have the potential to constrain the extended parameter space associated with the considered reheating scenarios. If these experiments detect positive signals, they would provide valuable constraints on the reheating mechanisms. For instance, if a signal is observed in the lower mass regime: $3.7\cdot 10^{-9 }~\text{eV} \lesssim m_a \lesssim 5.1\cdot 10^{-7} ~\text{eV}$ (cf. Table \ref{Tab:ma}), this would suggest that the reheating scenario via bosonic annihilation becomes less likely. Similarly, if a signal emerges in a larger mass range: $ 4.9 \cdot 10^{-5 }~\text{eV} \lesssim m_a \lesssim 2.5 \cdot 10^{-4} ~\text{eV}$ (cf. Table \ref{Tab:ma}), the reheating scenarios involving bosonic decay and annihilation with $n\lesssim 8$ could  be ruled out. This exclusion could be achieved without necessitating a fine-tuning of the misalignment angle.

	For ALPs, a portion of the parameter space can be probed by future Haloscope experiments, as indicated in Fig.~\ref{fig:gag}. However, it is important to note that the larger mass region remains outside the sensitivity of these experiments.
	Due to the distinct ordering of parameter space for different $n$ values, future experiments are particularly sensitive to scenarios with higher values of $n$. Consequently, if a signal is detected in the larger mass region, it could lead to the exclusion of scenarios with smaller values of $n$. Unlike the QCD axion case, the underlying type of inflaton-matter couplings cannot be readily deduced from these observations in the context of ALPs.

	%%%%%%%%%%%%%%%%%%%%%%%%%%%%%%%%%%%
	\section{Conclusions}\label{sec:conclusion}
	%%%%%%%%%%%%%%%%%%%%%%%%%%%%%%%%%%%
	In this work, we investigated the production of QCD axions and axion-like particles (ALPs) as candidates for dark matter (DM) via the vacuum misalignment mechanism {\it during} inflationary reheating. By assuming that the inflaton oscillates around a generic monomial potential $\sim \phi^n$, we derived the parameter space that gives rise to the correct relic abundance for reheating scenarios involving inflaton decay and annihilation. Additionally, we explored the experimental constraints by comparing the parameter space with current and future axion experiments.
	
	For QCD axions with $\Tosc > \Tqcd$, our analysis in Sec.~\ref{sec:QCD_case1} revealed a dependence of relic abundance on both the value of $n$ and the type of inflaton-matter couplings. In the contrasting scenario of $\Tosc < \Tqcd$, as elucidated in Sec.~\ref{sec:QCD_case2}, the relic abundance was shown to hinge solely on the value of $n$. Notably, we demonstrated that for reheating temperatures $4~\text{MeV} \lesssim \Trh \lesssim 1~\text{GeV}$, the parameter space capable of yielding the correct relic abundance can be significantly expanded compared to the conventional case where misalignment transpires in the radiation-dominated epoch \textit{after} reheating. This augmentation is  depicted in Fig.~\ref{fig:case2} and summarized in Table~\ref{Tab:ma}. For ALPs, we found that the parameter space can also be extended and is influenced only by the value of $n$.  We derived several general analytical expressions, such as Eq.~\eqref{eq:_rhoALP3} and Eq.~\eqref{eq:Omega_a}, which can reproduce the earlier  results~\cite{Arias:2021rer, Blinov:2019rhb} concerning  misalignment during early matter $(n=2)$  and  kination epochs $(n\to \infty)$.

	By further considering axion and ALP couplings to photons, we examined the constraints imposed by current and future experiments. Due to the dependence of  the enlarged parameter space  on the underlying reheating dynamics, the constraints differ for different reheating scenarios. For QCD axions, we found that current ADMX and CAPP experiments are already capable of ruling out certain parts of the expanded parameter space with $ 2 \cdot10^{-6}~\text{eV} \lesssim m_a\lesssim  5\cdot10^{-6}~\text{eV}$, particularly when $n\leq 4$ for both the decay and annihilation  scenarios. For $n=6$, we found that  ADMX and CAPP experiments exhibit the potential to probe $m_a \sim 5\cdot 10^{-6}~\text{eV}$ solely in the context of bosonic decay scenario. These results are shown in Fig.~\ref{fig:QCDgag}. For ALPs, our analysis identified Telescope experiments like MUSE are already capable  of constraining scenarios involving $n=8$ in the mass range of $2.7~\text{eV}-5.3~\text{eV}$, as depicted in Fig.~\ref{fig:gag}.
	
	By mapping the extended parameter space with future experimental projections, we found that forthcoming Haloscope experiments hold significant potential to constrain the parameter space associated with different reheating scenarios for both QCD axions and ALPs. We also highlighted that  positive detections at certain mass regimes could lead to exclusion of specific reheating scenarios. In particular, by assuming QCD axions as DM,  we found that the bosonic annihilation reheating scenarios become less likely if a signal is observed in the lower mass regime: $3.7\cdot 10^{-9 }~\text{eV} \lesssim m_a \lesssim 5.1\cdot 10^{-7} ~\text{eV}$. On the other hand, if a signal were to appear within a  range with larger mass: $ 4.9 \cdot 10^{-5 }~\text{eV} \lesssim m_a \lesssim 2.5 \cdot 10^{-4} ~\text{eV}$, the reheating scenarios involving bosonic decay and annihilation with $n \lesssim 8$ could be ruled out, unless the initial misalignment angle is not $\mathcal{O}(1)$. 
	%%%%%%%%%%%%%%%%%%%%%%%%%%%%%%%%%%%%%%%%%%%
	\section*{Acknowledgments}
	%%%%%%%%%%%%%%%%%%%%%%%%%%%%%%%%%%%%%%%%%%
	The author is grateful to Nicolás Bernal and Manuel Drees for  comments on the draft. YX receives support from the Cluster of Excellence ``Precision Physics, Fundamental Interactions, and Structure
	of Matter'' (PRISMA$^+$ EXC 2118/1) funded by the Deutsche Forschungsgemeinschaft (DFG, German Research Foundation) within the German Excellence Strategy (Project No. 39083149).
	\bibliographystyle{JHEP}
	\bibliography{biblio}

\providecommand{\href}[2]{#2}\begingroup\raggedright\begin{thebibliography}{100}

\bibitem{Preskill:1982cy}
J.~Preskill, M.B.~Wise and F.~Wilczek, \emph{{Cosmology of the Invisible
  Axion}}, \href{https://doi.org/10.1016/0370-2693(83)90637-8}{\emph{Phys.
  Lett. B} {\bfseries 120} (1983) 127}.

\bibitem{Abbott:1982af}
L.F.~Abbott and P.~Sikivie, \emph{{A Cosmological Bound on the Invisible
  Axion}}, \href{https://doi.org/10.1016/0370-2693(83)90638-X}{\emph{Phys.
  Lett. B} {\bfseries 120} (1983) 133}.

\bibitem{Dine:1982ah}
M.~Dine and W.~Fischler, \emph{{The Not So Harmless Axion}},
  \href{https://doi.org/10.1016/0370-2693(83)90639-1}{\emph{Phys. Lett. B}
  {\bfseries 120} (1983) 137}.

\bibitem{Arias:2012az}
P.~Arias, D.~Cadamuro, M.~Goodsell, J.~Jaeckel, J.~Redondo and A.~Ringwald,
  \emph{{WISPy Cold Dark Matter}},
  \href{https://doi.org/10.1088/1475-7516/2012/06/013}{\emph{JCAP} {\bfseries
  06} (2012) 013} [\href{https://arxiv.org/abs/1201.5902}{{\ttfamily
  1201.5902}}].

\bibitem{Peccei:1977hh}
R.D.~Peccei and H.R.~Quinn, \emph{{CP Conservation in the Presence of
  Instantons}}, \href{https://doi.org/10.1103/PhysRevLett.38.1440}{\emph{Phys.
  Rev. Lett.} {\bfseries 38} (1977) 1440}.

\bibitem{Peccei:1977ur}
R.D.~Peccei and H.R.~Quinn, \emph{{Constraints Imposed by CP Conservation in
  the Presence of Instantons}},
  \href{https://doi.org/10.1103/PhysRevD.16.1791}{\emph{Phys. Rev. D}
  {\bfseries 16} (1977) 1791}.

\bibitem{Weinberg:1977ma}
S.~Weinberg, \emph{{A New Light Boson?}},
  \href{https://doi.org/10.1103/PhysRevLett.40.223}{\emph{Phys. Rev. Lett.}
  {\bfseries 40} (1978) 223}.

\bibitem{Wilczek:1977pj}
F.~Wilczek, \emph{{Problem of Strong $P$ and $T$ Invariance in the Presence of
  Instantons}}, \href{https://doi.org/10.1103/PhysRevLett.40.279}{\emph{Phys.
  Rev. Lett.} {\bfseries 40} (1978) 279}.

\bibitem{Conlon:2006tq}
J.P.~Conlon, \emph{{The QCD axion and moduli stabilisation}},
  \href{https://doi.org/10.1088/1126-6708/2006/05/078}{\emph{JHEP} {\bfseries
  05} (2006) 078} [\href{https://arxiv.org/abs/hep-th/0602233}{{\ttfamily
  hep-th/0602233}}].

\bibitem{Svrcek:2006yi}
P.~Svrcek and E.~Witten, \emph{{Axions In String Theory}},
  \href{https://doi.org/10.1088/1126-6708/2006/06/051}{\emph{JHEP} {\bfseries
  06} (2006) 051} [\href{https://arxiv.org/abs/hep-th/0605206}{{\ttfamily
  hep-th/0605206}}].

\bibitem{Arvanitaki:2009fg}
A.~Arvanitaki, S.~Dimopoulos, S.~Dubovsky, N.~Kaloper and J.~March-Russell,
  \emph{{String Axiverse}},
  \href{https://doi.org/10.1103/PhysRevD.81.123530}{\emph{Phys. Rev. D}
  {\bfseries 81} (2010) 123530}
  [\href{https://arxiv.org/abs/0905.4720}{{\ttfamily 0905.4720}}].

\bibitem{Acharya:2010zx}
B.S.~Acharya, K.~Bobkov and P.~Kumar, \emph{{An M Theory Solution to the Strong
  CP Problem and Constraints on the Axiverse}},
  \href{https://doi.org/10.1007/JHEP11(2010)105}{\emph{JHEP} {\bfseries 11}
  (2010) 105} [\href{https://arxiv.org/abs/1004.5138}{{\ttfamily 1004.5138}}].

\bibitem{Cicoli:2012sz}
M.~Cicoli, M.~Goodsell and A.~Ringwald, \emph{{The type IIB string axiverse and
  its low-energy phenomenology}},
  \href{https://doi.org/10.1007/JHEP10(2012)146}{\emph{JHEP} {\bfseries 10}
  (2012) 146} [\href{https://arxiv.org/abs/1206.0819}{{\ttfamily 1206.0819}}].

\bibitem{Choi:2020rgn}
K.~Choi, S.H.~Im and C.~Sub~Shin, \emph{{Recent Progress in the Physics of
  Axions and Axion-Like Particles}},
  \href{https://doi.org/10.1146/annurev-nucl-120720-031147}{\emph{Ann. Rev.
  Nucl. Part. Sci.} {\bfseries 71} (2021) 225}
  [\href{https://arxiv.org/abs/2012.05029}{{\ttfamily 2012.05029}}].

\bibitem{Agrawal:2022yvu}
P.~Agrawal, K.V.~Berghaus, J.~Fan, A.~Hook, G.~Marques-Tavares and T.~Rudelius,
  \emph{{Some open questions in axion theory}},  in \emph{{Snowmass 2021}}, 3,
  2022 [\href{https://arxiv.org/abs/2203.08026}{{\ttfamily 2203.08026}}].

\bibitem{Sikivie:2006ni}
P.~Sikivie, \emph{{Axion Cosmology}},
  \href{https://doi.org/10.1007/978-3-540-73518-2_2}{\emph{Lect. Notes Phys.}
  {\bfseries 741} (2008) 19}
  [\href{https://arxiv.org/abs/astro-ph/0610440}{{\ttfamily
  astro-ph/0610440}}].

\bibitem{Bernal:2021yyb}
N.~Bernal, F.~Hajkarim and Y.~Xu, \emph{{Axion Dark Matter in the Time of
  Primordial Black Holes}},
  \href{https://doi.org/10.1103/PhysRevD.104.075007}{\emph{Phys. Rev. D}
  {\bfseries 104} (2021) 075007}
  [\href{https://arxiv.org/abs/2107.13575}{{\ttfamily 2107.13575}}].

\bibitem{Schiavone:2021imu}
F.~Schiavone, D.~Montanino, A.~Mirizzi and F.~Capozzi, \emph{{Axion-like
  particles from primordial black holes shining through the Universe}},
  \href{https://doi.org/10.1088/1475-7516/2021/08/063}{\emph{JCAP} {\bfseries
  08} (2021) 063} [\href{https://arxiv.org/abs/2107.03420}{{\ttfamily
  2107.03420}}].

\bibitem{Bernal:2021bbv}
N.~Bernal, Y.F.~Perez-Gonzalez, Y.~Xu and O.~Zapata, \emph{{ALP dark matter in
  a primordial black hole dominated universe}},
  \href{https://doi.org/10.1103/PhysRevD.104.123536}{\emph{Phys. Rev. D}
  {\bfseries 104} (2021) 123536}
  [\href{https://arxiv.org/abs/2110.04312}{{\ttfamily 2110.04312}}].

\bibitem{Mazde:2022sdx}
K.~Mazde and L.~Visinelli, \emph{{The interplay between the dark matter axion
  and primordial black holes}},
  \href{https://doi.org/10.1088/1475-7516/2023/01/021}{\emph{JCAP} {\bfseries
  01} (2023) 021} [\href{https://arxiv.org/abs/2209.14307}{{\ttfamily
  2209.14307}}].

\bibitem{Turner:1989vc}
M.S.~Turner, \emph{{Windows on the Axion}},
  \href{https://doi.org/10.1016/0370-1573(90)90172-X}{\emph{Phys. Rept.}
  {\bfseries 197} (1990) 67}.

\bibitem{Marsh:2015xka}
D.J.E.~Marsh, \emph{{Axion Cosmology}},
  \href{https://doi.org/10.1016/j.physrep.2016.06.005}{\emph{Phys. Rept.}
  {\bfseries 643} (2016) 1} [\href{https://arxiv.org/abs/1510.07633}{{\ttfamily
  1510.07633}}].

\bibitem{DiLuzio:2020wdo}
L.~Di~Luzio, M.~Giannotti, E.~Nardi and L.~Visinelli, \emph{{The landscape of
  QCD axion models}},
  \href{https://doi.org/10.1016/j.physrep.2020.06.002}{\emph{Phys. Rept.}
  {\bfseries 870} (2020) 1} [\href{https://arxiv.org/abs/2003.01100}{{\ttfamily
  2003.01100}}].

\bibitem{Sikivie:2020zpn}
P.~Sikivie, \emph{{Invisible Axion Search Methods}},
  \href{https://doi.org/10.1103/RevModPhys.93.015004}{\emph{Rev. Mod. Phys.}
  {\bfseries 93} (2021) 015004}
  [\href{https://arxiv.org/abs/2003.02206}{{\ttfamily 2003.02206}}].

\bibitem{Kolb:1990vq}
E.W.~Kolb and M.S.~Turner, \emph{{The Early Universe}}, vol.~69 (1990),
  \href{https://doi.org/10.1201/9780429492860}{10.1201/9780429492860}.

\bibitem{Takahashi:2018tdu}
F.~Takahashi, W.~Yin and A.H.~Guth, \emph{{QCD axion window and low-scale
  inflation}}, \href{https://doi.org/10.1103/PhysRevD.98.015042}{\emph{Phys.
  Rev. D} {\bfseries 98} (2018) 015042}
  [\href{https://arxiv.org/abs/1805.08763}{{\ttfamily 1805.08763}}].

\bibitem{Graham:2018jyp}
P.W.~Graham and A.~Scherlis, \emph{{Stochastic axion scenario}},
  \href{https://doi.org/10.1103/PhysRevD.98.035017}{\emph{Phys. Rev. D}
  {\bfseries 98} (2018) 035017}
  [\href{https://arxiv.org/abs/1805.07362}{{\ttfamily 1805.07362}}].

\bibitem{Co:2019jts}
R.T.~Co, L.J.~Hall and K.~Harigaya, \emph{{Axion Kinetic Misalignment
  Mechanism}},
  \href{https://doi.org/10.1103/PhysRevLett.124.251802}{\emph{Phys. Rev. Lett.}
  {\bfseries 124} (2020) 251802}
  [\href{https://arxiv.org/abs/1910.14152}{{\ttfamily 1910.14152}}].

\bibitem{Chang:2019tvx}
C.-F.~Chang and Y.~Cui, \emph{{New Perspectives on Axion Misalignment
  Mechanism}}, \href{https://doi.org/10.1103/PhysRevD.102.015003}{\emph{Phys.
  Rev. D} {\bfseries 102} (2020) 015003}
  [\href{https://arxiv.org/abs/1911.11885}{{\ttfamily 1911.11885}}].

\bibitem{Barman:2021rdr}
B.~Barman, N.~Bernal, N.~Ramberg and L.~Visinelli, \emph{{QCD Axion Kinetic
  Misalignment without Prejudice}},
  \href{https://doi.org/10.3390/universe8120634}{\emph{Universe} {\bfseries 8}
  (2022) 634} [\href{https://arxiv.org/abs/2111.03677}{{\ttfamily
  2111.03677}}].

\bibitem{Steinhardt:1983ia}
P.J.~Steinhardt and M.S.~Turner, \emph{{Saving the Invisible Axion}},
  \href{https://doi.org/10.1016/0370-2693(83)90727-X}{\emph{Phys. Lett. B}
  {\bfseries 129} (1983) 51}.

\bibitem{Lazarides:1990xp}
G.~Lazarides, R.K.~Schaefer, D.~Seckel and Q.~Shafi, \emph{{Dilution of
  Cosmological Axions by Entropy Production}},
  \href{https://doi.org/10.1016/0550-3213(90)90244-8}{\emph{Nucl. Phys. B}
  {\bfseries 346} (1990) 193}.

\bibitem{Kawasaki:1995vt}
M.~Kawasaki, T.~Moroi and T.~Yanagida, \emph{{Can decaying particles raise the
  upper bound on the Peccei-Quinn scale?}},
  \href{https://doi.org/10.1016/0370-2693(96)00743-5}{\emph{Phys. Lett. B}
  {\bfseries 383} (1996) 313}
  [\href{https://arxiv.org/abs/hep-ph/9510461}{{\ttfamily hep-ph/9510461}}].

\bibitem{Giudice:2000ex}
G.F.~Giudice, E.W.~Kolb and A.~Riotto, \emph{{Largest temperature of the
  radiation era and its cosmological implications}},
  \href{https://doi.org/10.1103/PhysRevD.64.023508}{\emph{Phys. Rev. D}
  {\bfseries 64} (2001) 023508}
  [\href{https://arxiv.org/abs/hep-ph/0005123}{{\ttfamily hep-ph/0005123}}].

\bibitem{Grin:2007yg}
D.~Grin, T.L.~Smith and M.~Kamionkowski, \emph{{Axion constraints in
  non-standard thermal histories}},
  \href{https://doi.org/10.1103/PhysRevD.77.085020}{\emph{Phys. Rev. D}
  {\bfseries 77} (2008) 085020}
  [\href{https://arxiv.org/abs/0711.1352}{{\ttfamily 0711.1352}}].

\bibitem{Visinelli:2009kt}
L.~Visinelli and P.~Gondolo, \emph{{Axion cold dark matter in non-standard
  cosmologies}}, \href{https://doi.org/10.1103/PhysRevD.81.063508}{\emph{Phys.
  Rev. D} {\bfseries 81} (2010) 063508}
  [\href{https://arxiv.org/abs/0912.0015}{{\ttfamily 0912.0015}}].

\bibitem{Nelson:2018via}
A.E.~Nelson and H.~Xiao, \emph{{Axion Cosmology with Early Matter Domination}},
  \href{https://doi.org/10.1103/PhysRevD.98.063516}{\emph{Phys. Rev. D}
  {\bfseries 98} (2018) 063516}
  [\href{https://arxiv.org/abs/1807.07176}{{\ttfamily 1807.07176}}].

\bibitem{Visinelli:2018wza}
L.~Visinelli and J.~Redondo, \emph{{Axion Miniclusters in Modified Cosmological
  Histories}}, \href{https://doi.org/10.1103/PhysRevD.101.023008}{\emph{Phys.
  Rev. D} {\bfseries 101} (2020) 023008}
  [\href{https://arxiv.org/abs/1808.01879}{{\ttfamily 1808.01879}}].

\bibitem{Ramberg:2019dgi}
N.~Ramberg and L.~Visinelli, \emph{{Probing the Early Universe with Axion
  Physics and Gravitational Waves}},
  \href{https://doi.org/10.1103/PhysRevD.99.123513}{\emph{Phys. Rev. D}
  {\bfseries 99} (2019) 123513}
  [\href{https://arxiv.org/abs/1904.05707}{{\ttfamily 1904.05707}}].

\bibitem{Blinov:2019rhb}
N.~Blinov, M.J.~Dolan, P.~Draper and J.~Kozaczuk, \emph{{Dark matter targets
  for axionlike particle searches}},
  \href{https://doi.org/10.1103/PhysRevD.100.015049}{\emph{Phys. Rev. D}
  {\bfseries 100} (2019) 015049}
  [\href{https://arxiv.org/abs/1905.06952}{{\ttfamily 1905.06952}}].

\bibitem{Blinov:2019jqc}
N.~Blinov, M.J.~Dolan and P.~Draper, \emph{{Imprints of the Early Universe on
  Axion Dark Matter Substructure}},
  \href{https://doi.org/10.1103/PhysRevD.101.035002}{\emph{Phys. Rev. D}
  {\bfseries 101} (2020) 035002}
  [\href{https://arxiv.org/abs/1911.07853}{{\ttfamily 1911.07853}}].

\bibitem{Arias:2021rer}
P.~Arias, N.~Bernal, D.~Karamitros, C.~Maldonado, L.~Roszkowski and M.~Venegas,
  \emph{{New opportunities for axion dark matter searches in nonstandard
  cosmological models}},
  \href{https://doi.org/10.1088/1475-7516/2021/11/003}{\emph{JCAP} {\bfseries
  11} (2021) 003} [\href{https://arxiv.org/abs/2107.13588}{{\ttfamily
  2107.13588}}].

\bibitem{Arias:2022qjt}
P.~Arias, N.~Bernal, J.K.~Osi\'nski and L.~Roszkowski, \emph{{Dark matter
  axions in the early universe with a period of increasing temperature}},
  \href{https://doi.org/10.1088/1475-7516/2023/05/028}{\emph{JCAP} {\bfseries
  05} (2023) 028} [\href{https://arxiv.org/abs/2207.07677}{{\ttfamily
  2207.07677}}].

\bibitem{Drees:2017iod}
M.~Drees and F.~Hajkarim, \emph{{Dark Matter Production in an Early Matter
  Dominated Era}},
  \href{https://doi.org/10.1088/1475-7516/2018/02/057}{\emph{JCAP} {\bfseries
  02} (2018) 057} [\href{https://arxiv.org/abs/1711.05007}{{\ttfamily
  1711.05007}}].

\bibitem{Bernal:2018qlk}
N.~Bernal, M.~Dutra, Y.~Mambrini, K.~Olive, M.~Peloso and M.~Pierre,
  \emph{{Spin-2 Portal Dark Matter}},
  \href{https://doi.org/10.1103/PhysRevD.97.115020}{\emph{Phys. Rev. D}
  {\bfseries 97} (2018) 115020}
  [\href{https://arxiv.org/abs/1803.01866}{{\ttfamily 1803.01866}}].

\bibitem{Maity:2018dgy}
D.~Maity and P.~Saha, \emph{{Connecting CMB anisotropy and cold dark matter
  phenomenology via reheating}},
  \href{https://doi.org/10.1103/PhysRevD.98.103525}{\emph{Phys. Rev. D}
  {\bfseries 98} (2018) 103525}
  [\href{https://arxiv.org/abs/1801.03059}{{\ttfamily 1801.03059}}].

\bibitem{Maity:2018exj}
D.~Maity and P.~Saha, \emph{{CMB constraints on dark matter phenomenology via
  reheating in Minimal plateau inflation}},
  \href{https://doi.org/10.1016/j.dark.2019.100317}{\emph{Phys. Dark Univ.}
  {\bfseries 25} (2019) 100317}
  [\href{https://arxiv.org/abs/1804.10115}{{\ttfamily 1804.10115}}].

\bibitem{Bernal:2019mhf}
N.~Bernal, F.~Elahi, C.~Maldonado and J.~Unwin, \emph{{Ultraviolet Freeze-in
  and Non-Standard Cosmologies}},
  \href{https://doi.org/10.1088/1475-7516/2019/11/026}{\emph{JCAP} {\bfseries
  11} (2019) 026} [\href{https://arxiv.org/abs/1909.07992}{{\ttfamily
  1909.07992}}].

\bibitem{Garcia:2020eof}
M.A.G.~Garcia, K.~Kaneta, Y.~Mambrini and K.A.~Olive, \emph{{Reheating and
  Post-inflationary Production of Dark Matter}},
  \href{https://doi.org/10.1103/PhysRevD.101.123507}{\emph{Phys. Rev. D}
  {\bfseries 101} (2020) 123507}
  [\href{https://arxiv.org/abs/2004.08404}{{\ttfamily 2004.08404}}].

\bibitem{Bernal:2021qrl}
N.~Bernal and Y.~Xu, \emph{{Polynomial inflation and dark matter}},
  \href{https://doi.org/10.1140/epjc/s10052-021-09694-5}{\emph{Eur. Phys. J. C}
  {\bfseries 81} (2021) 877}
  [\href{https://arxiv.org/abs/2106.03950}{{\ttfamily 2106.03950}}].

\bibitem{Calibbi:2021fld}
L.~Calibbi, F.~D'Eramo, S.~Junius, L.~Lopez-Honorez and A.~Mariotti,
  \emph{{Displaced new physics at colliders and the early universe before its
  first second}}, \href{https://doi.org/10.1007/JHEP05(2021)234}{\emph{JHEP}
  {\bfseries 05} (2021) 234}
  [\href{https://arxiv.org/abs/2102.06221}{{\ttfamily 2102.06221}}].

\bibitem{Ahmed:2022tfm}
A.~Ahmed, B.~Grzadkowski and A.~Socha, \emph{{Higgs boson induced reheating and
  ultraviolet frozen-in dark matter}},
  \href{https://doi.org/10.1007/JHEP02(2023)196}{\emph{JHEP} {\bfseries 02}
  (2023) 196} [\href{https://arxiv.org/abs/2207.11218}{{\ttfamily
  2207.11218}}].

\bibitem{Barman:2022tzk}
B.~Barman, N.~Bernal, Y.~Xu and O.~Zapata, \emph{{Ultraviolet freeze-in with a
  time-dependent inflaton decay}},
  \href{https://doi.org/10.1088/1475-7516/2022/07/019}{\emph{JCAP} {\bfseries
  07} (2022) 019} [\href{https://arxiv.org/abs/2202.12906}{{\ttfamily
  2202.12906}}].

\bibitem{Banerjee:2022fiw}
A.~Banerjee and D.~Chowdhury, \emph{{Fingerprints of freeze-in dark matter in
  an early matter-dominated era}},
  \href{https://doi.org/10.21468/SciPostPhys.13.2.022}{\emph{SciPost Phys.}
  {\bfseries 13} (2022) 022}
  [\href{https://arxiv.org/abs/2204.03670}{{\ttfamily 2204.03670}}].

\bibitem{Bernal:2022wck}
N.~Bernal and Y.~Xu, \emph{{WIMPs during reheating}},
  \href{https://doi.org/10.1088/1475-7516/2022/12/017}{\emph{JCAP} {\bfseries
  12} (2022) 017} [\href{https://arxiv.org/abs/2209.07546}{{\ttfamily
  2209.07546}}].

\bibitem{Bhattiprolu:2022sdd}
P.N.~Bhattiprolu, G.~Elor, R.~McGehee and A.~Pierce, \emph{{Freezing-in
  hadrophilic dark matter at low reheating temperatures}},
  \href{https://doi.org/10.1007/JHEP01(2023)128}{\emph{JHEP} {\bfseries 01}
  (2023) 128} [\href{https://arxiv.org/abs/2210.15653}{{\ttfamily
  2210.15653}}].

\bibitem{Haque:2023yra}
M.R.~Haque, D.~Maity and R.~Mondal, \emph{{WIMPs, FIMPs, and Inflaton
  phenomenology via reheating, CMB and $\Delta N_{eff}$}},
  \href{https://arxiv.org/abs/2301.01641}{{\ttfamily 2301.01641}}.

\bibitem{Chowdhury:2023jft}
D.~Chowdhury and A.~Hait, \emph{{Thermalization in the presence of a
  time-dependent dissipation and its impact on dark matter production}},
  \href{https://arxiv.org/abs/2302.06654}{{\ttfamily 2302.06654}}.

\bibitem{Silva-Malpartida:2023yks}
J.~Silva-Malpartida, N.~Bernal, J.~Jones-P\'erez and R.A.~Lineros, \emph{{From
  WIMPs to FIMPs with Low Reheating Temperatures}},
  \href{https://arxiv.org/abs/2306.14943}{{\ttfamily 2306.14943}}.

\bibitem{Becker:2023tvd}
M.~Becker, E.~Copello, J.~Harz, J.~Lang and Y.~Xu, \emph{{Confronting Dark
  Matter Freeze-In during Reheating with Constraints from Inflation}},
  \href{https://arxiv.org/abs/2306.17238}{{\ttfamily 2306.17238}}.

\bibitem{Gan:2023jbs}
X.~Gan and Y.-D.~Tsai, \emph{{Cosmic Millicharge Background and Reheating
  Probes}},  \href{https://arxiv.org/abs/2308.07951}{{\ttfamily 2308.07951}}.

\bibitem{Irastorza:2018dyq}
I.G.~Irastorza and J.~Redondo, \emph{{New experimental approaches in the search
  for axion-like particles}},
  \href{https://doi.org/10.1016/j.ppnp.2018.05.003}{\emph{Prog. Part. Nucl.
  Phys.} {\bfseries 102} (2018) 89}
  [\href{https://arxiv.org/abs/1801.08127}{{\ttfamily 1801.08127}}].

\bibitem{Adams:2022pbo}
C.B.~Adams et~al., \emph{{Axion Dark Matter}},  in \emph{{Snowmass 2021}}, 3,
  2022 [\href{https://arxiv.org/abs/2203.14923}{{\ttfamily 2203.14923}}].

\bibitem{Garcia:2020wiy}
M.A.G.~Garcia, K.~Kaneta, Y.~Mambrini and K.A.~Olive, \emph{{Inflaton
  Oscillations and Post-Inflationary Reheating}},
  \href{https://doi.org/10.1088/1475-7516/2021/04/012}{\emph{JCAP} {\bfseries
  04} (2021) 012} [\href{https://arxiv.org/abs/2012.10756}{{\ttfamily
  2012.10756}}].

\bibitem{Turner:1983he}
M.S.~Turner, \emph{{Coherent Scalar Field Oscillations in an Expanding
  Universe}}, \href{https://doi.org/10.1103/PhysRevD.28.1243}{\emph{Phys. Rev.
  D} {\bfseries 28} (1983) 1243}.

\bibitem{Allahverdi:2010xz}
R.~Allahverdi, R.~Brandenberger, F.-Y.~Cyr-Racine and A.~Mazumdar,
  \emph{{Reheating in Inflationary Cosmology: Theory and Applications}},
  \href{https://doi.org/10.1146/annurev.nucl.012809.104511}{\emph{Ann. Rev.
  Nucl. Part. Sci.} {\bfseries 60} (2010) 27}
  [\href{https://arxiv.org/abs/1001.2600}{{\ttfamily 1001.2600}}].

\bibitem{Amin:2014eta}
M.A.~Amin, M.P.~Hertzberg, D.I.~Kaiser and J.~Karouby, \emph{{Nonperturbative
  Dynamics Of Reheating After Inflation: A Review}},
  \href{https://doi.org/10.1142/S0218271815300037}{\emph{Int. J. Mod. Phys. D}
  {\bfseries 24} (2014) 1530003}
  [\href{https://arxiv.org/abs/1410.3808}{{\ttfamily 1410.3808}}].

\bibitem{Lozanov:2019jxc}
K.D.~Lozanov, \emph{{Lectures on Reheating after Inflation}},
  \href{https://arxiv.org/abs/1907.04402}{{\ttfamily 1907.04402}}.

\bibitem{Kallosh:2013hoa}
R.~Kallosh and A.~Linde, \emph{{Universality Class in Conformal Inflation}},
  \href{https://doi.org/10.1088/1475-7516/2013/07/002}{\emph{JCAP} {\bfseries
  07} (2013) 002} [\href{https://arxiv.org/abs/1306.5220}{{\ttfamily
  1306.5220}}].

\bibitem{Starobinsky:1980te}
A.A.~Starobinsky, \emph{{A New Type of Isotropic Cosmological Models Without
  Singularity}},
  \href{https://doi.org/10.1016/0370-2693(80)90670-X}{\emph{Phys. Lett. B}
  {\bfseries 91} (1980) 99}.

\bibitem{Drees:2021wgd}
M.~Drees and Y.~Xu, \emph{{Small field polynomial inflation: reheating,
  radiative stability and lower bound}},
  \href{https://doi.org/10.1088/1475-7516/2021/09/012}{\emph{JCAP} {\bfseries
  09} (2021) 012} [\href{https://arxiv.org/abs/2104.03977}{{\ttfamily
  2104.03977}}].

\bibitem{Drees:2022aea}
M.~Drees and Y.~Xu, \emph{{Large field polynomial inflation: parameter space,
  predictions and (double) eternal nature}},
  \href{https://doi.org/10.1088/1475-7516/2022/12/005}{\emph{JCAP} {\bfseries
  12} (2022) 005} [\href{https://arxiv.org/abs/2209.07545}{{\ttfamily
  2209.07545}}].

\bibitem{Xu:2022qpx}
Y.~Xu, \emph{{Polynomial Inflation and Its Aftermath}}, Ph.D. thesis, U. Bonn
  (main), 2022.

\bibitem{Ichikawa:2008ne}
K.~Ichikawa, T.~Suyama, T.~Takahashi and M.~Yamaguchi, \emph{{Primordial
  Curvature Fluctuation and Its Non-Gaussianity in Models with Modulated
  Reheating}}, \href{https://doi.org/10.1103/PhysRevD.78.063545}{\emph{Phys.
  Rev. D} {\bfseries 78} (2008) 063545}
  [\href{https://arxiv.org/abs/0807.3988}{{\ttfamily 0807.3988}}].

\bibitem{Lozanov:2016hid}
K.D.~Lozanov and M.A.~Amin, \emph{{Equation of State and Duration to Radiation
  Domination after Inflation}},
  \href{https://doi.org/10.1103/PhysRevLett.119.061301}{\emph{Phys. Rev. Lett.}
  {\bfseries 119} (2017) 061301}
  [\href{https://arxiv.org/abs/1608.01213}{{\ttfamily 1608.01213}}].

\bibitem{Haque:2022kez}
M.R.~Haque and D.~Maity, \emph{{Gravitational reheating}},
  \href{https://doi.org/10.1103/PhysRevD.107.043531}{\emph{Phys. Rev. D}
  {\bfseries 107} (2023) 043531}
  [\href{https://arxiv.org/abs/2201.02348}{{\ttfamily 2201.02348}}].

\bibitem{Dufaux:2006ee}
J.F.~Dufaux, G.N.~Felder, L.~Kofman, M.~Peloso and D.~Podolsky,
  \emph{{Preheating with trilinear interactions: Tachyonic resonance}},
  \href{https://doi.org/10.1088/1475-7516/2006/07/006}{\emph{JCAP} {\bfseries
  07} (2006) 006} [\href{https://arxiv.org/abs/hep-ph/0602144}{{\ttfamily
  hep-ph/0602144}}].

\bibitem{Maity:2018qhi}
D.~Maity and P.~Saha, \emph{{(P)reheating after minimal Plateau Inflation and
  constraints from CMB}},
  \href{https://doi.org/10.1088/1475-7516/2019/07/018}{\emph{JCAP} {\bfseries
  07} (2019) 018} [\href{https://arxiv.org/abs/1811.11173}{{\ttfamily
  1811.11173}}].

\bibitem{Saha:2020bis}
P.~Saha, S.~Anand and L.~Sriramkumar, \emph{{Accounting for the time evolution
  of the equation of state parameter during reheating}},
  \href{https://doi.org/10.1103/PhysRevD.102.103511}{\emph{Phys. Rev. D}
  {\bfseries 102} (2020) 103511}
  [\href{https://arxiv.org/abs/2005.01874}{{\ttfamily 2005.01874}}].

\bibitem{Peloso:2000hy}
M.~Peloso and L.~Sorbo, \emph{{Preheating of massive fermions after inflation:
  Analytical results}},
  \href{https://doi.org/10.1088/1126-6708/2000/05/016}{\emph{JHEP} {\bfseries
  05} (2000) 016} [\href{https://arxiv.org/abs/hep-ph/0003045}{{\ttfamily
  hep-ph/0003045}}].

\bibitem{Co:2022bgh}
R.T.~Co, Y.~Mambrini and K.A.~Olive, \emph{{Inflationary gravitational
  leptogenesis}},
  \href{https://doi.org/10.1103/PhysRevD.106.075006}{\emph{Phys. Rev. D}
  {\bfseries 106} (2022) 075006}
  [\href{https://arxiv.org/abs/2205.01689}{{\ttfamily 2205.01689}}].

\bibitem{Barman:2022qgt}
B.~Barman, S.~Cl\'ery, R.T.~Co, Y.~Mambrini and K.A.~Olive, \emph{{Gravity as a
  portal to reheating, leptogenesis and dark matter}},
  \href{https://doi.org/10.1007/JHEP12(2022)072}{\emph{JHEP} {\bfseries 12}
  (2022) 072} [\href{https://arxiv.org/abs/2210.05716}{{\ttfamily
  2210.05716}}].

\bibitem{GrillidiCortona:2015jxo}
G.~Grilli~di Cortona, E.~Hardy, J.~Pardo~Vega and G.~Villadoro, \emph{{The QCD
  axion, precisely}},
  \href{https://doi.org/10.1007/JHEP01(2016)034}{\emph{JHEP} {\bfseries 01}
  (2016) 034} [\href{https://arxiv.org/abs/1511.02867}{{\ttfamily
  1511.02867}}].

\bibitem{Borsanyi:2016ksw}
S.~Borsanyi et~al., \emph{{Calculation of the axion mass based on
  high-temperature lattice quantum chromodynamics}},
  \href{https://doi.org/10.1038/nature20115}{\emph{Nature} {\bfseries 539}
  (2016) 69} [\href{https://arxiv.org/abs/1606.07494}{{\ttfamily 1606.07494}}].

\bibitem{ParticleDataGroup:2022pth}
{\scshape Particle Data Group} collaboration, \emph{{Review of Particle
  Physics}}, \href{https://doi.org/10.1093/ptep/ptac097}{\emph{PTEP} {\bfseries
  2022} (2022) 083C01}.

\bibitem{Kawasaki:2000en}
M.~Kawasaki, K.~Kohri and N.~Sugiyama, \emph{{MeV scale reheating temperature
  and thermalization of neutrino background}},
  \href{https://doi.org/10.1103/PhysRevD.62.023506}{\emph{Phys. Rev. D}
  {\bfseries 62} (2000) 023506}
  [\href{https://arxiv.org/abs/astro-ph/0002127}{{\ttfamily
  astro-ph/0002127}}].

\bibitem{Hannestad:2004px}
S.~Hannestad, \emph{{What is the lowest possible reheating temperature?}},
  \href{https://doi.org/10.1103/PhysRevD.70.043506}{\emph{Phys. Rev. D}
  {\bfseries 70} (2004) 043506}
  [\href{https://arxiv.org/abs/astro-ph/0403291}{{\ttfamily
  astro-ph/0403291}}].

\bibitem{Graham:2015ouw}
P.W.~Graham, I.G.~Irastorza, S.K.~Lamoreaux, A.~Lindner and K.A.~van Bibber,
  \emph{{Experimental Searches for the Axion and Axion-Like Particles}},
  \href{https://doi.org/10.1146/annurev-nucl-102014-022120}{\emph{Ann. Rev.
  Nucl. Part. Sci.} {\bfseries 65} (2015) 485}
  [\href{https://arxiv.org/abs/1602.00039}{{\ttfamily 1602.00039}}].

\bibitem{ADMX:2018gho}
{\scshape ADMX} collaboration, \emph{{A Search for Invisible Axion Dark Matter
  with the Axion Dark Matter Experiment}},
  \href{https://doi.org/10.1103/PhysRevLett.120.151301}{\emph{Phys. Rev. Lett.}
  {\bfseries 120} (2018) 151301}
  [\href{https://arxiv.org/abs/1804.05750}{{\ttfamily 1804.05750}}].

\bibitem{ADMX:2019uok}
{\scshape ADMX} collaboration, \emph{{Extended Search for the Invisible Axion
  with the Axion Dark Matter Experiment}},
  \href{https://doi.org/10.1103/PhysRevLett.124.101303}{\emph{Phys. Rev. Lett.}
  {\bfseries 124} (2020) 101303}
  [\href{https://arxiv.org/abs/1910.08638}{{\ttfamily 1910.08638}}].

\bibitem{Crisosto:2019fcj}
N.~Crisosto, P.~Sikivie, N.S.~Sullivan, D.B.~Tanner, J.~Yang and G.~Rybka,
  \emph{{ADMX SLIC: Results from a Superconducting $LC$ Circuit Investigating
  Cold Axions}},
  \href{https://doi.org/10.1103/PhysRevLett.124.241101}{\emph{Phys. Rev. Lett.}
  {\bfseries 124} (2020) 241101}
  [\href{https://arxiv.org/abs/1911.05772}{{\ttfamily 1911.05772}}].

\bibitem{ADMX:2021nhd}
{\scshape ADMX} collaboration, \emph{{Search for Invisible Axion Dark Matter in
  the 3.3\textendash{}4.2\,\,\ensuremath{\mu}eV Mass Range}},
  \href{https://doi.org/10.1103/PhysRevLett.127.261803}{\emph{Phys. Rev. Lett.}
  {\bfseries 127} (2021) 261803}
  [\href{https://arxiv.org/abs/2110.06096}{{\ttfamily 2110.06096}}].

\bibitem{ADMX:2021mio}
{\scshape ADMX} collaboration, \emph{{Dark matter axion search using a
  Josephson Traveling wave parametric amplifier}},
  \href{https://doi.org/10.1063/5.0122907}{\emph{Rev. Sci. Instrum.} {\bfseries
  94} (2023) 044703} [\href{https://arxiv.org/abs/2110.10262}{{\ttfamily
  2110.10262}}].

\bibitem{Lee:2020cfj}
S.~Lee, S.~Ahn, J.~Choi, B.R.~Ko and Y.K.~Semertzidis, \emph{{Axion Dark Matter
  Search around 6.7 $\mu$eV}},
  \href{https://doi.org/10.1103/PhysRevLett.124.101802}{\emph{Phys. Rev. Lett.}
  {\bfseries 124} (2020) 101802}
  [\href{https://arxiv.org/abs/2001.05102}{{\ttfamily 2001.05102}}].

\bibitem{Jeong:2020cwz}
J.~Jeong, S.~Youn, S.~Bae, J.~Kim, T.~Seong, J.E.~Kim et~al., \emph{{Search for
  Invisible Axion Dark Matter with a Multiple-Cell Haloscope}},
  \href{https://doi.org/10.1103/PhysRevLett.125.221302}{\emph{Phys. Rev. Lett.}
  {\bfseries 125} (2020) 221302}
  [\href{https://arxiv.org/abs/2008.10141}{{\ttfamily 2008.10141}}].

\bibitem{CAPP:2020utb}
{\scshape CAPP} collaboration, \emph{{First Results from an Axion Haloscope at
  CAPP around 10.7 $\mu$eV}},
  \href{https://doi.org/10.1103/PhysRevLett.126.191802}{\emph{Phys. Rev. Lett.}
  {\bfseries 126} (2021) 191802}
  [\href{https://arxiv.org/abs/2012.10764}{{\ttfamily 2012.10764}}].

\bibitem{Lee:2022mnc}
Y.~Lee, B.~Yang, H.~Yoon, M.~Ahn, H.~Park, B.~Min et~al., \emph{{Searching for
  Invisible Axion Dark Matter with an 18~T Magnet Haloscope}},
  \href{https://doi.org/10.1103/PhysRevLett.128.241805}{\emph{Phys. Rev. Lett.}
  {\bfseries 128} (2022) 241805}
  [\href{https://arxiv.org/abs/2206.08845}{{\ttfamily 2206.08845}}].

\bibitem{Kim:2022hmg}
J.~Kim et~al., \emph{{Near-Quantum-Noise Axion Dark Matter Search at CAPP
  around 9.5\,\,\ensuremath{\mu}eV}},
  \href{https://doi.org/10.1103/PhysRevLett.130.091602}{\emph{Phys. Rev. Lett.}
  {\bfseries 130} (2023) 091602}
  [\href{https://arxiv.org/abs/2207.13597}{{\ttfamily 2207.13597}}].

\bibitem{Todarello:2023hdk}
E.~Todarello, M.~Regis, J.~Reynoso-Cordova, M.~Taoso, D.~Vaz, J.~Brinchmann
  et~al., \emph{{Robust bounds on ALP dark matter from dwarf spheroidal
  galaxies in the optical MUSE-Faint survey}},
  \href{https://arxiv.org/abs/2307.07403}{{\ttfamily 2307.07403}}.

\bibitem{DMRadio:2022pkf}
{\scshape DMRadio} collaboration, \emph{{Projected sensitivity of DMRadio-m3: A
  search for the QCD axion below 1\,\,\ensuremath{\mu}eV}},
  \href{https://doi.org/10.1103/PhysRevD.106.103008}{\emph{Phys. Rev. D}
  {\bfseries 106} (2022) 103008}
  [\href{https://arxiv.org/abs/2204.13781}{{\ttfamily 2204.13781}}].

\bibitem{Alesini:2019nzq}
D.~Alesini et~al., \emph{{KLASH Conceptual Design Report}},
  \href{https://arxiv.org/abs/1911.02427}{{\ttfamily 1911.02427}}.

\bibitem{Diaz-Morcillo:2021psa}
A.~D\'\i{}az-Morcillo et~al., \emph{{Design of New Resonant Haloscopes in the
  Search for the Dark Matter Axion: A Review of the First Steps in the RADES
  Collaboration}},
  \href{https://doi.org/10.3390/universe8010005}{\emph{Universe} {\bfseries 8}
  (2021) 5} [\href{https://arxiv.org/abs/2111.14510}{{\ttfamily 2111.14510}}].

\bibitem{Stern:2016bbw}
I.~Stern, \emph{{ADMX Status}},
  \href{https://doi.org/10.22323/1.282.0198}{\emph{PoS} {\bfseries ICHEP2016}
  (2016) 198} [\href{https://arxiv.org/abs/1612.08296}{{\ttfamily
  1612.08296}}].

\bibitem{DeMiguel:2023nmz}
J.~De~Miguel and J.F.~Hern\'andez-Cabrera, \emph{{Discovery prospects with the
  Dark-photons \& Axion-Like particles Interferometer--part I}},
  \href{https://arxiv.org/abs/2303.03997}{{\ttfamily 2303.03997}}.

\bibitem{Lawson:2019brd}
M.~Lawson, A.J.~Millar, M.~Pancaldi, E.~Vitagliano and F.~Wilczek,
  \emph{{Tunable axion plasma haloscopes}},
  \href{https://doi.org/10.1103/PhysRevLett.123.141802}{\emph{Phys. Rev. Lett.}
  {\bfseries 123} (2019) 141802}
  [\href{https://arxiv.org/abs/1904.11872}{{\ttfamily 1904.11872}}].

\bibitem{Beurthey:2020yuq}
S.~Beurthey et~al., \emph{{MADMAX Status Report}},
  \href{https://arxiv.org/abs/2003.10894}{{\ttfamily 2003.10894}}.

\bibitem{McAllister:2017lkb}
B.T.~McAllister, G.~Flower, J.~Kruger, E.N.~Ivanov, M.~Goryachev, J.~Bourhill
  et~al., \emph{{The ORGAN Experiment: An axion haloscope above 15 GHz}},
  \href{https://doi.org/10.1016/j.dark.2017.09.010}{\emph{Phys. Dark Univ.}
  {\bfseries 18} (2017) 67} [\href{https://arxiv.org/abs/1706.00209}{{\ttfamily
  1706.00209}}].

\bibitem{Aja:2022csb}
B.~Aja et~al., \emph{{The Canfranc Axion Detection Experiment (CADEx): search
  for axions at 90 GHz with Kinetic Inductance Detectors}},
  \href{https://doi.org/10.1088/1475-7516/2022/11/044}{\emph{JCAP} {\bfseries
  11} (2022) 044} [\href{https://arxiv.org/abs/2206.02980}{{\ttfamily
  2206.02980}}].

\bibitem{BREAD:2021tpx}
{\scshape BREAD} collaboration, \emph{{Broadband Solenoidal Haloscope for
  Terahertz Axion Detection}},
  \href{https://doi.org/10.1103/PhysRevLett.128.131801}{\emph{Phys. Rev. Lett.}
  {\bfseries 128} (2022) 131801}
  [\href{https://arxiv.org/abs/2111.12103}{{\ttfamily 2111.12103}}].

\bibitem{Baryakhtar:2018doz}
M.~Baryakhtar, J.~Huang and R.~Lasenby, \emph{{Axion and hidden photon dark
  matter detection with multilayer optical haloscopes}},
  \href{https://doi.org/10.1103/PhysRevD.98.035006}{\emph{Phys. Rev. D}
  {\bfseries 98} (2018) 035006}
  [\href{https://arxiv.org/abs/1803.11455}{{\ttfamily 1803.11455}}].

\bibitem{AxionLimits}
C.~O'Hare, ``cajohare/axionlimits: Axionlimits.''
  \url{https://cajohare.github.io/AxionLimits/}, July, 2020.
\newblock 10.5281/zenodo.3932430.

\end{thebibliography}\endgroup
	
\end{document}